\documentclass[twocolumn]{aastex701}

\usepackage{amsmath,amssymb,amsfonts,amsthm}
\usepackage{graphicx}
\usepackage{bm}
\usepackage{algorithm}
\usepackage{algorithmic}
\usepackage{booktabs}
\usepackage{xcolor}
\definecolor{aspred}{RGB}{180,30,45}

\newtheorem{proposition}{Proposition}

\newcommand{\vect}[1]{\bm{#1}}
\newcommand{\mat}[1]{\mathbf{#1}}

\newcommand{\diag}{\mathrm{diag}}
\newcommand{\softplus}{\sigma}
\newcommand{\dd}{\mathrm{d}}
\newcommand{\reals}{\mathbb{R}}

\newcommand{\one}{\mathbf{1}}

\newcommand{\nd}[1]{\parbox[t]{0.62\linewidth}{#1\strut}}

\shorttitle{NestyNet. I. Accurate Surrogates and Analytic Derivatives}
\shortauthors{Ibata et al.}

\begin{document}

\title{NestyNet. I. Physics Functions Are Hard to Fit with Neural Networks:\\
A Framework for Accurate Surrogates and Analytic Derivatives}

\author[0000-0002-3292-9709]{Rodrigo Ibata}
\affiliation{Universit\'e de Strasbourg, CNRS, Observatoire astronomique de Strasbourg, UMR 7550, F-67000 Strasbourg, France}
\email[show]{rodrigo.ibata@astro.unistra.fr}

\author[0000-0001-8392-3836]{Wassim Tenachi}
\affiliation{Mila - Quebec Artificial Intelligence Institute}
\affiliation{D\'epartement de physique, Universit\'e de Montr\'eal}
\affiliation{Ciela - Montreal Institute for Astrophysical Data Analysis and Machine Learning}
\email{wassim.tenachi@umontreal.ca}

\author[0000-0002-8788-8174]{Foivos Diakogiannis}
\affiliation{Technology, Commonwealth Scientific and Industrial Research Organisation (CSIRO), Kensington, WA 6151, Australia}
\email{Foivos.Diakogiannis@data61.csiro.au}

\author[0000-0003-1559-1053]{Neil Ibata}
\affiliation{Department of Human Evolutionary Biology, Harvard University, Cambridge, MA, USA}
\email{neilibata@fas.harvard.edu}

\author[0009-0008-7455-1880]{Anirudh Shankar}
\affiliation{Universit\'e de Strasbourg, CNRS, Observatoire astronomique de Strasbourg, UMR 7550, F-67000 Strasbourg, France}
\email{anirudh.shankar@astro.unistra.fr}

\begin{abstract}
  Many of the smooth functions that matter most in physics are precisely the
  ones that standard neural network methods struggle to fit accurately.
  Here we present NestyNet, a coupled model-and-optimizer framework capable
  of fitting such targets to high accuracy while also delivering their gradients,
  Hessians, Laplacians, and antiderivatives analytically and at low cost.
  This makes it a natural substrate for scientific machine learning tasks.
  The model is a
  deterministic segmented analytic surrogate, and its optimizer is a
  second-order Levenberg--Marquardt scheme whose damping and linear solves
  are tailored to the stiff, strongly correlated parameter geometries
  induced by multiscale and sharply structured targets typical in physics
  and other scientific applications.  
  
  On the AI Feynman benchmark of 120
  physics equations, NestyNet achieves median improvement factors of
  $2\,100\times$ for function values, $1\,400\times$ for first
  derivatives, and $780\times$ for second derivatives relative to standard neural networks
  trained with first-order optimization (Adam).  Even
  after refining those fits with quasi-Newton (L-BFGS) optimization, 
  the corresponding improvements are $540\times$, $450\times$, and $250\times$.
  Owing to the analytic design it is up to $\approx 44\times$
  faster than vectorized automatic-differentiation (autograd) baselines,
  with a margin growing with model size. The same
  analytic-derivative framework also supports vector- and complex-valued
  targets, measurement uncertainties in both inputs and outputs, and
  constraints, and its modules can be composed flexibly to build
  scientifically useful model architectures, all without reverting to
  autograd. Together,
  these components provide a practical modular framework for fitting
  difficult scientific surrogates while delivering accurate differential
  operators for subsequent analysis.
\end{abstract}


\section{Introduction}
\label{sec:introduction}

The functions that arise in physics are generally not easy to fit or
interpolate.  They are often multiscale, highly anisotropic, or sharply
structured while remaining smooth, as in boundary layers, resonant
responses, phase transitions, or nearly piecewise regimes smoothed by finite
resolution or dissipation.

In such settings the difficulty in fitting neural surrogates is not one of
representational power, but one of \emph{parameter-space geometry}.
Even a shallow network is in principle a universal
approximator~\citep{Cybenko1989,Hornik1989}, so the right parameters exist.
The difficulty is that a poorly conditioned landscape makes them hard to
find.  This is the problem we target.

The elementary brick of our
architecture is a shallow network whose geometry we can control.  Such bricks
compose, by summing and by nesting, into surrogates whose derivatives remain
analytic and whose block structure is preserved.  The optimizer is second
order, built to exploit that structure as it moves through the landscapes
these compositions produce.  Standard multi-layer perceptrons (MLPs) trained with first-order
optimizers perform remarkably well when the effective parameter directions
are only weakly correlated.  But for many scientific targets a faithful
representation induces strongly correlated,
sloppy~\citep[in the sense of parameter-sensitivity eigenvalues spanning many
decades;][]{Transtrum2011}, and stiff parameter directions,
carving narrow curved valleys into the loss landscape that in turn lead to
slow or fragile convergence under first-order updates.  Automatic
differentiation (hereafter autograd) computes derivatives of the resulting
network exactly, but
if the fitted representation itself is poorly conditioned or inefficient,
then the corresponding differential operators inherit those deficiencies.

Building machine learning models for physics is especially challenging because we usually
require more than good pointwise fits to function values.  For instance, in
physics-informed neural networks (PINNs, \citealp{Raissi2019,Karniadakis2021})
derivative errors directly corrupt the field equations being enforced.  In
trajectory-based dynamics and inverse problems errors destabilize integration
and bias inferred forces, in Hamiltonian dynamics inaccurate
$\partial H/\partial p$ and $\partial H/\partial q$ produce secular energy
drift, and downstream structure-sensitive tasks such as stability analysis
and symbolic regression (the subject of Papers III and IV) rely on mixed
partials that noisy derivatives corrupt from the outset.

These failures are all tied to parameter-space geometry.  When the target
induces strongly correlated parameter directions, a first-order optimizer
can fit the function in a weak sense while representing local curvature
poorly.  We therefore require a surrogate that resolves the difficult
geometry of the target and an optimizer that moves efficiently in the
resulting stiff parameter space.

The present research program developed organically from earlier work aiming
to learn a neural net approximation to the generating function for the
canonical transformation from positions and velocities of stars along orbits
in the Milky Way to their action-angle coordinate
equivalents~\citep{Ibata2021ActionFinder}.  The accuracy of the derived
transformations in that work was hampered by the use of first-order
optimizers and deep networks with ReLU (rectified linear unit)
activations, which led us to consider
second-order alternatives and networks whose derivative structure could be
more easily controlled.

We address this challenge by designing a coupled model-and-optimizer
construct that we call NestyNet, assembling a family of composable,
analytically differentiable surrogate models for scientific computing.  The
model side is a segmented softplus-activated architecture in which the basic
neural network block is a sum of smooth hinge elements.  This representation
is well matched to the sharply structured but still differentiable response
surfaces common in physics, and it yields closed-form expressions for
gradients, Hessians, and Laplacians.  The optimizer is a
Levenberg--Marquardt (LM;~\citealt{Levenberg1944,Marquardt1963}) scheme tailored
to the segment structure through a particular ordering, 
which exposes a useful decomposition of the
Gauss--Newton system.  Our design is predictive, in that it looks ahead
along the optimization path, proposing more aggressive parameter updates
when advantageous.  We also show how quasi-Newton
L-BFGS~\citep{Nocedal1980,LiuNocedal1989} steps can be harvested at
essentially no additional cost and be used to automatically steer LM out of
locally stagnant regions of the optimization.

The model and optimizer combination enables several advances developed in
detail below.  First, gradients, Hessians, and Laplacians are available
analytically rather than by traversal of the network's computational
graph, so once the
surrogate is fit, the associated differential operators are available in
closed form.  Second, for softplus segmented models we introduce an optional seed-free
initialization that embeds the best affine least-squares fit exactly inside
the segmented softplus basis, then
adds segments deterministically by residual-aligned projection, yielding
repeatable fits and an importance-ordered segment sequence that makes it
easy to find the complexity a given fit requires.  Third, the
Gauss--Newton matrix $\mat{J}^\top\mat{J}$ (the Gram matrix of the
Jacobian columns) decomposes exactly into segment-diagonal
and cross-segment overlap terms, which we package as segment-native
preconditioning linear algebra (SPLA; \S\ref{sec:spla}), a segment-aware framework that treats a
small subspace exactly and verifies the resulting step against
the true objective, letting the second-order solves scale to large models.
Its theory and constructions are developed in \citet[hereafter the SPLA theory paper]{NestyNet2026Ia}.  Fourth, quadratic path
extrapolation and Woodbury-based fast paths~\citep{Woodbury1950} accelerate the Levenberg--Marquardt
optimization, each in the phase of the solve where it is effective
(\S\ref{sec:predictive}).  Fifth, we release a
composable adaptor ecosystem (\S\ref{sec:adaptors}) that turns NestyNet into
a general scientific-optimization backend, covering vector- and
complex-valued surrogates, partial differential equation (PDE) and dynamics
residuals, and probability densities.  The present paper validates its core,
the surrogate and the optimizer.  In short,
NestyNet is designed for the physically important regime in which the
fitting problem itself is hard.

The present work establishes the deterministic NestyNet substrate.
In \citet[hereafter Paper~II]{NestyNet2026b}~we construct its coherent function-space
posterior, while in~\citet[hereafter Paper~III]{NestyNet2026c} and~\citet[hereafter Paper~IV]{NestyNet2026d}
we use its analytic differential structure for symbolic-regression and
differential-equation discovery.  The
contribution of the present paper is therefore deliberately foundational, a
surrogate whose local differential structure is trustworthy and cheap enough
to enable those downstream discovery tasks.

This work is organized as follows.  Section~\ref{sec:segmented} introduces
the segmented architecture and \S\ref{sec:derivatives} its analytic
derivative and antiderivative machinery.  We then turn to the SPLA
preconditioning framework in \S\ref{sec:spla}, followed by the predictive
Levenberg--Marquardt optimizer in \S\ref{sec:predictive}.
Section~\ref{sec:adaptors} sketches the composable adaptor framework that
ties the pieces together.  Section~\ref{sec:applications} demonstrates the framework
as an emulator of PDE solutions, and \S\ref{sec:benchmarks}
presents numerical experiments, including the AI Feynman benchmark.
Section~\ref{sec:cbe_application} applies the resulting derivative accuracy to
an astrophysical inverse problem, the vertical acceleration of the Galactic
disk.  We conclude in
\S\ref{sec:conclusions}.  Table~\ref{tab:notation} collects the
principal notation, grouped by the section in which each symbol is
introduced.

\begin{table*}[!t]
\centering
\small
\caption{Principal notation, grouped by the section of first use.
Deliberate, locally scoped reuses of a symbol, and near-collisions between
similar symbols, are noted in square brackets.}
\label{tab:notation}
\begin{minipage}[t]{0.49\textwidth}
\begin{tabular}{@{}l l@{}}
\toprule
Symbol & \nd{Represents} \\
\midrule
\multicolumn{2}{@{}l}{\emph{Segmented networks (\S\ref{sec:segmented})}}\\
$N_x$, $N_{\rm out}$ & \nd{input and output dimensionalities} \\
$S$, $W$ & \nd{segments per model; hinge channels per segment} \\
$a_{os}$, $b_{os}$, $K_{os}$ & \nd{hinge amplitude, bias, and input weights (kernel)} \\
$z_{os}$ & \nd{hinge preactivation $K_{os}\!\cdot\!x+b_{os}$} \\
$\softplus(z)$ & \nd{softplus activation $\log(1+e^{z})$} \\
\midrule
\multicolumn{2}{@{}l}{\emph{Analytic derivatives (\S\ref{sec:derivatives})}}\\
$g_{os}$ & \nd{sigmoid gate factor $\softplus'(z_{os})$} \\
$\operatorname{Li}_{\nu}$ & \nd{polylogarithm of order $\nu$} \\
$A(z)$ & \nd{softplus antiderivative $-\operatorname{Li}_{2}(-e^{z})$} \\
\midrule
\multicolumn{2}{@{}l}{\emph{Preconditioner (\S\ref{sec:spla})}}\\
$\mat{J}$, $r$, $h$ & \nd{residual Jacobian, residual vector, parameter step} \\
$\lambda$, $\mat{D}_{\rm lm}$ & \nd{LM damping; damping scaling diagonal} \\
$g$ & \nd{objective gradient [bare $g$; distinct from the gate $g_{os}$]} \\
$\mat{A}_{\lambda}$ & \nd{damped curvature $\mat{J}^{\top}\mat{J}+\lambda\mat{D}_{\rm lm}$} \\
$\mat{B}_{\lambda}$, $\mat{E}$ & \nd{within-segment block base; cross-segment coupling} \\
$\mat{Z}$ & \nd{coarse space of directions treated exactly [reused in \S\ref{sec:predictive} for the local-manifold basis]} \\
$\widehat{\mat{A}}_{\lambda}$ & \nd{damping-whitened operator} \\
$d_{\rm eff}(\lambda)$ & \nd{effective dimension of the whitened operator} \\
$\phi(\theta)$ & \nd{least-squares objective at parameter vector $\theta$} \\
\bottomrule
\end{tabular}
\end{minipage}\hfill
\begin{minipage}[t]{0.49\textwidth}
\begin{tabular}{@{}l l@{}}
\toprule
Symbol & \nd{Represents} \\
\midrule
\multicolumn{2}{@{}l}{\emph{Predictive LM (\S\ref{sec:predictive})}}\\
$\mat{W}$ & \nd{diagonal data-weight matrix (identity if unweighted)} \\
$\mat{H}$ & \nd{Gauss--Newton matrix $\mat{J}^{\top}\mat{W}\mat{J}$} \\
$\mat{S}$ & \nd{solved-geometry scaling diagonal [bold; the italic $S$ is the segment count]} \\
$\rho$, $\widehat{\Delta\phi}$ & \nd{trust ratio; predicted objective decrease} \\
$z(\theta)$, $\Sigma$ & \nd{standardized residual $\Sigma^{-1/2}r$ [distinct from the height $z$]; data covariance} \\
$\eta_f$, $\mat{Q}$, $\mat{P}(t)$ & \nd{patch strengths; patched data precision; prior precision} \\
$\mat{U}$ & \nd{curvature-factor columns of the direct predictor (Woodbury factors)} \\
$\tau(\lambda)$ & \nd{damping-aware predictor acceptance threshold} \\
\midrule
\multicolumn{2}{@{}l}{\emph{Benchmarks (\S\ref{sec:benchmarks})}}\\
$e$ & \nd{fit error $\hat f-f$ of surrogate $\hat f$ to target $f$} \\
$\Vert\nabla e\Vert_{L^2}$, $|e|_{H^2}$ & \nd{gradient and curvature (Frobenius--Hessian) error norms} \\
$\rho_e$ & \nd{spectral tightness ratio [distinct from the trust ratio $\rho$]} \\
$k_e$ & \nd{characteristic error wavenumber $\Vert\nabla e\Vert_{L^2}/\Vert e\Vert_{L^2}$} \\
\midrule
\multicolumn{2}{@{}l}{\emph{Astrophysical application (\S\ref{sec:cbe_application})}}\\
$f(z,v)$, $\hat g$ & \nd{phase-space distribution function; fitted $\ln\hat f$ ($z$: height, $v$: vertical velocity)} \\
$K_z(z)$ & \nd{vertical acceleration [distinct from the kernels $K_{os}$]} \\
$\sigma$, $a$ & \nd{slab velocity dispersion and scale height [this section only]} \\
\bottomrule
\end{tabular}
\end{minipage}
\end{table*}

\section{Segmented Neural Networks}
\label{sec:segmented}

\subsection{Architecture}

The foundation of NestyNet is a shallow, single-hidden-layer segmented neural
network architecture, in which the output is a sum over $S$ independent ``segments'':
\begin{equation}
f_o(x) = \sum_{s=1}^{S} a_{os} \, \softplus\!\left( z_{os} \right)
\label{eq:segmented}
\end{equation}
where we define
\begin{equation}
z_{os} \equiv \sum_{j=1}^{N_x} K_{osj} x_j + b_{os},
\qquad
\softplus(z) \equiv \log(1 + e^z),
\label{eq:zos_def}
\end{equation}
$j \in \{1, \ldots, N_x\}$ indexes input dimensions, $o \in \{1, \ldots, N_{\rm out}\}$ indexes output dimensions, and each
segment $s$ has parameters $(a_{os}, b_{os}, K_{osj})$.  Vector-valued
targets are thus native to the architecture: each output carries its own
segment parameters, all outputs are fitted jointly against the stacked
residual, and the adaptors of \S\ref{sec:adaptors} couple outputs
where shared structure is required.

The implementation supports a slightly more general ``segment width''
$W$, in which each segment contains $W$ parallel soft-hinges per output:
\begin{align}
f_o(x) &= \sum_{s=1}^{S}\sum_{w=1}^{W} a_{osw}\,
\softplus\!\left( z_{osw} \right) \,,\\
z_{osw} &\equiv \sum_{j=1}^{N_x} K_{oswj} x_j + b_{osw} \,.
\label{eq:segmented_width}
\end{align}
We present the $W{=}1$ form~\eqref{eq:segmented} for clarity, but $W$ is
handled explicitly in the optimizer. The preconditioner machinery of
\S\ref{sec:spla} and the SPLA theory paper treats
each width channel as its own atomic hinge (one output--segment
soft-hinge of the segmented model), while gate-based diagnostics collapse
a segment's $W$ channels by a Euclidean norm over $w$.  In
practice, $W$ acts as a small intra-segment basis that improves expressivity
while preserving the analytic derivative structure, with all derivatives remaining
closed-form and efficient.

The softplus acts as a smooth hinge (vanishing for $z \ll 0$ and
approaching $z$ for $z \gg 0$), so each segment activates as the input
crosses the hyperplane defined by $K_{os}$ and $b_{os}$.  The network
approximates functions by summing soft hinges with learned amplitudes, as
ReLU networks do with piecewise-linear pieces, but with the smoothness
needed for well-defined Hessians and Laplacians.

The released library provides a suite of model families sharing this
analytic-derivative contract: weight-normalized and complex-valued variants
of Eq.~\eqref{eq:segmented}, trigonometric units for periodic targets, and
normalized Gaussian-mixture and multivariate-$t$ densities.  This paper
concentrates on the softplus form, the most versatile workhorse of the
family.

\subsection{Initialization and Optimizer-Facing Structure}

We provide a deterministic
initialization procedure (which we call ``canonical'') that embeds the best affine fit
within the same segmented basis.  The affine initialization makes use of
the identity $\softplus(z)-\softplus(-z)=z$: for each output, the fitted linear
form is represented by an opposed pair of hinges with opposite preactivations
and opposite amplitudes, and the intercept by a zero-slope hinge of value
$\softplus(0)=\log 2$.  Additional hinges are then grown by
residual-aligned projections, typically producing an importance-ordered segment
sequence and low initial overlap.  Parameters are provided to the optimizer in
a deliberate order (Output-Segment-Type-Parameter) so that same-output blocks are contiguous, which is what later makes
the SPLA algebra (\S\ref{sec:spla}) transparent.  Incremental segment growth and optional pruning-and-reseeding of collinear
hinges improve stability in difficult fits. The main point for this paper
is that canonical initialization produces repeatable, low-overlap starting
conditions favorable for second-order fitting.

This logic is managed by a \emph{gauge controller}, which applies the
canonical greedy initialization on the \emph{live coordinates} (those with
non-negligible empirical variance) and is reused when dormant segments are
activated later in training, so the initial fit and incremental growth
share the deterministic policy.

For a stacked two-stage surrogate $f(x)=f_1(f_0(x))$ the first stage's
target is not observed directly, so the canonical construction is extended
in four deterministic steps.  The inner stage starts as an exact affine
lift, $x\mapsto[x,0,\ldots,0]$, and the outer stage is canonically
initialized on the live hidden coordinates.  Targets for the extra hidden
channels come from the fit itself.  We pull the output residual $r_y$ back
to hidden space as $u=\mat J_h^\top\Sigma^{-1}r_y$, where
$\mat J_h=\partial f_1/\partial h$ and $\Sigma$ is the data covariance, and
then remove the part of $u$ that the inputs already explain linearly.  What
remains is the deterministic target for those channels.  The remaining inner segments are fitted to this enriched target,
and the outer stage is re-initialized on the result.  The idea extends
recursively to deeper stacks.

\section{Analytic Derivatives and Antiderivatives}
\label{sec:derivatives}

A key advantage of the segmented architecture is that all relevant
derivatives (value, gradient, Hessian, Hessian contractions, and
Laplacian) admit closed-form expressions, and so, as we show in
\S\ref{sec:antiderivatives}, do antiderivatives along arbitrary line
segments.  In contrast, autograd finds these by applying the
chain rule to the unreduced computational graph.  Our closed-form route
collapses the segment structure analytically, reuses the gate factors,
and exposes low-rank and trace contractions such as $\nabla^2 f$ and
$v^{\top}Hv$ at much lower graph and memory cost.

\subsection{Gradients, Hessians, and Laplacians}
\label{sec:derivatives_down}

For the segmented architecture~\eqref{eq:segmented}, with pre-activation
$z_{os}$ as in~\eqref{eq:zos_def}, we define the gate factor
\begin{equation}
g_{os}(x) = \softplus'(z_{os}) = \frac{1}{1 + e^{-z_{os}}} \in (0,1),
\label{eq:gate_factor}
\end{equation}
which is the sigmoid.  The function value is the segment sum,
\begin{equation}
f_o(x) = \sum_{s=1}^S a_{os} \, \softplus(z_{os}),
\end{equation}
and the gradient and Hessian with respect to the inputs take the closed forms
\begin{equation}
\frac{\partial f_o}{\partial x_j} = \sum_s a_{os} \, g_{os} \, K_{osj},
\label{eq:gradient}
\end{equation}
\begin{equation}
\frac{\partial^2 f_o}{\partial x_i \partial x_j} = \sum_s a_{os} \, g_{os}(1-g_{os}) \, K_{osi} K_{osj}.
\label{eq:hessian}
\end{equation}
The factor $g(1-g)$, the derivative of the sigmoid, peaks at $g = 1/2$
(gate half-open) and vanishes as $g \to 0$ or $g \to 1$, which naturally
suppresses Hessian contributions from saturated segments.

For PDE applications the Laplacian $\nabla^2 f_o$ is required at every
collocation point (a sample point at which the PDE residual is enforced).
A naive evaluation via the full Hessian~\eqref{eq:hessian}
would cost $\mathcal{O}(N_x^2)$ per output per sample, but the Laplacian needs only
the trace, which collapses the inner sum:
\begin{equation}
\nabla^2 f_o
= \sum_j \frac{\partial^2 f_o}{\partial x_j^2}
= \sum_s a_{os} \, g_{os}(1-g_{os}) \, \|K_{os}\|_2^2,
\label{eq:laplacian}
\end{equation}
with $\|K_{os}\|_2^2 = \sum_j K_{osj}^2$ the squared Euclidean norm of the
kernel.  This is the Laplacian in the Euclidean input metric.  Curvilinear 
or transformed coordinates introduce the usual metric factors.  
Once $\|K_{os}\|_2^2$ has been precomputed and the gate factors
$g_{os}(1-g_{os})$ are available, evaluating the Laplacian costs
$\mathcal{O}(S)$ per sample and output. In the width-$W$ model
this becomes $\mathcal{O}(SW)$. Including the preactivation
computation from scratch, the cost is $\mathcal{O}(SWN_x)$,
whereas constructing the full Hessian would cost
$\mathcal{O}(SWN_x^2)$.

\subsection{Closed-Form Antiderivatives: The Polylogarithm Ladder}
\label{sec:antiderivatives}

The derivative formulas above are the descending half of a single
structure.  For the polylogarithm
$\operatorname{Li}_\nu(w)=\sum_{k\ge1}w^{k}/k^{\nu}$~\citep{Lewin1981} one
has $\frac{\dd}{\dd z}\operatorname{Li}_{\nu}(-e^{z})
=\operatorname{Li}_{\nu-1}(-e^{z})$ at every order $\nu$, and the softplus
atom sits on this ladder:
\begin{align}
\softplus(z)&=-\operatorname{Li}_{1}(-e^{z}),
\qquad
g=-\operatorname{Li}_{0}(-e^{z}),
\nonumber\\
g(1-g)&=-\operatorname{Li}_{-1}(-e^{z}).
\label{eq:polylog_ladder_down}
\end{align}
Differentiation moves \emph{down} the ladder into the elementary rational
functions of the gate that appear in
Eqs.~\eqref{eq:gradient}--\eqref{eq:laplacian}, while integration moves
\emph{up} into the classical polylogarithms:
\begin{equation}
A(z):=\int^{z}\!\softplus(u)\,\dd u=-\operatorname{Li}_{2}(-e^{z}),
\label{eq:polylog_ladder_up}
\end{equation}
with one further integration giving $-\operatorname{Li}_{3}(-e^{z})$, and
so on.  Every antiderivative of the segmented model therefore retains
the segment structure and the preactivations $z_{os}$ unchanged, with only
the gate kernel moved one rung up the ladder.

Because every preactivation is affine in the input, the restriction of the
model to any straight line $x(t)=x_{0}+t\,v$ has $z_{os}(t)$ affine in $t$
with slope $\alpha_{os}=K_{os}\!\cdot\!v$, so exact line integrals cost one
antiderivative evaluation per segment endpoint:
\begin{equation}
\int_{t_1}^{t_2}\! f_o(x_{0}+t\,v)\,\dd t
=\sum_{s}\frac{a_{os}}{\alpha_{os}}
\Bigl[A\bigl(z_{os}(t_2)\bigr)-A\bigl(z_{os}(t_1)\bigr)\Bigr],
\label{eq:line_integral}
\end{equation}
with the exact limit $a_{os}\,\softplus(z_{os})\,(t_{2}-t_{1})$ for
degenerate slopes $\alpha_{os}\to0$.  Axis-parallel integrals, cumulative
profiles, and fluxes or circulations along polygonal paths are special
cases, at cost $\mathcal{O}(SW)$ per integral.
Iterating the ladder yields exact integrals over hyperrectangles
$R=\prod_{j}[l_{j},u_{j}]$ by inclusion--exclusion over the $2^{N_x}$
corners $x_{\varepsilon}$ of $R$ (with $(x_{\varepsilon})_j=u_j$ when
$\varepsilon_j=1$ and $l_j$ when $\varepsilon_j=0$),
\begin{align}
\int_{R} f_o\,\dd^{N_x}x
=\sum_{s}\frac{a_{os}}{\prod_{j}K_{osj}}
&\sum_{\varepsilon\in\{0,1\}^{N_x}}
(-1)^{N_x-|\varepsilon|}\nonumber\\
&\times
\bigl[-\operatorname{Li}_{N_x+1}\bigl(-e^{z_{os}(x_{\varepsilon})}\bigr)\bigr],
\label{eq:box_integral}
\end{align}
an exact cubature at $\mathcal{O}(SW\,2^{N_x})$ polylogarithm evaluations
(degenerate axes reduce to lower rungs as above).

The mirror identity $\softplus(z)-\softplus(-z)=z$ that underpins the
canonical initialization (\S\ref{sec:segmented}) also lifts up the
ladder. Integrating it once gives
\begin{equation}
A(z)+A(-z)=\frac{z^{2}}{2}+\frac{\pi^{2}}{6},
\label{eq:integrated_mirror}
\end{equation}
the classical dilogarithm inversion formula.  It guarantees that the
affine channel embedded by the canonical initialization integrates exactly
to the quadratic it represents, and it also makes the numerical evaluation
stable, with $A$ computed by series only for $z\le0$, where the argument
$-e^{z}$ lies in $(-1,0]$, and reflected through
Eq.~\eqref{eq:integrated_mirror} otherwise.  The higher rungs needed by
Eq.~\eqref{eq:box_integral} are evaluated the same way, with an
accelerated alternating series~\citep{CohenVillegasZagier2000} on the
convergent side and the
general polylogarithm inversion formula on the other.  The released
implementation evaluates
these kernels to $\approx10^{-15}$ relative accuracy in double precision
at every order we have tested (up to $\operatorname{Li}_{11}$, i.e.\
$N_x=10$).

Closed-form integrability is not generally preserved under nonlinear
composition, so the exact polylogarithm antiderivatives apply directly to the
single-stage segmented representation, while stacked compositions
(\S\ref{sec:segmented}) are integrated by quadrature of their cheap
analytic evaluations.  The integration step therefore adds no numerical
error of its own. For a single-layer
32-segment surrogate of problem \#1 of the AI Feynman physics-equation
suite~\citep[][the benchmark used throughout \S\ref{sec:benchmarks}]{Udrescu2020},
fitted in two seconds
with predictive LM, the closed-form axis, line, and box integrals agree
with adaptive quadrature of the same surrogate to $10^{-13}$--$10^{-16}$.

\section{A Segment-Native Preconditioner for Levenberg--Marquardt}
\label{sec:spla}

Each Levenberg--Marquardt (LM) step solves the damped normal equations
\begin{equation}
(\mat{J}^\top\mat{J} + \lambda \mat{D}_{\rm lm}) \, h = -\mat{J}^\top r ,
\label{eq:lm}
\end{equation}
with $h$ the parameter step, $\mat{J}$ the Jacobian, $r$ the residual,
$\lambda$ the damping, and $\mat{D}_{\rm lm}$ a diagonal scaling.  For $n$
parameters a direct factorization of~\eqref{eq:lm} costs $\mathcal{O}(n^3)$.
The more scalable alternative is preconditioned conjugate gradient
(PCG)~\citep{Saad2003}, which converges at a rate set by the preconditioned
condition number $\kappa$.  PCG is therefore only as good as its preconditioner,
which motivated the one developed here, targeted at segmented models.

However, SPLA was not needed for the results of this paper or of
Papers~II--IV, whose problems are
small enough for standard direct factorization.
Section~\ref{sec:spla_performance} previews its payoff at larger parameter
counts.

In the systems our fits produce, what sets $\kappa$ is the geometry of the
soft hinges.  Two hinges with
nearly the same orientation and offset produce almost the same output, so the
fit can raise one amplitude and lower the other with little change in the
prediction.  Through the softplus mirror identity
$\softplus(z)-\softplus(-z)=z$, a nearly \emph{opposed} pair does the same up to
a term linear in the inputs.  Each such near-coincidence is a soft, nearly flat
direction of the least-squares landscape, a sloppy mode of the kind ubiquitous
in multiparameter physics models.  A handful of
these soft modes sitting among the stiff, well-determined directions is exactly
what makes the Gauss--Newton curvature ill-conditioned, and it is why the
second-order step that must invert that curvature is the bottleneck of the fit.
A diagonal (Jacobi) preconditioner cannot remove them, because the offending
coupling is between the parameters of different hinges.

By design, the segmented architecture also supplies the cure.  Because every parameter belongs to a
single segment, the curvature splits exactly into cheap within-segment
blocks plus a cross-segment coupling that carries the soft modes.  The dominant
soft directions can be written down in closed form from the hinge parameters,
with no eigensolve, or captured spectrally
by a randomized sketch~\citep{Halko2011,FrangellaTroppUdell2023} of the
damping-whitened operator (a probe with a
small set of random vectors, whose responses span the operator's dominant
eigendirections), which proves to be the main workhorse.  
The coupling $\mat{E}$ is not itself low rank.  What is often, though
not always, numerically compressible is the part of it that harms the solve,
and the SPLA theory paper characterizes when.  Where it is, SPLA treats that small
subspace exactly and leaves an ordinary block-preconditioned
PCG to handle a well-conditioned remainder, escalating to block or
Schur treatment~\citep{Zhang2005} when it is not.  A single small
projected solve certifies the chosen subspace.  The remainder of this section
records the main features of the framework and the solver ladder it unifies.

At a fixed linearization and damping the solver faces one symmetric
positive-definite system, $\mat{A}_{\lambda}h=-g$, where
$\mat{A}_{\lambda}=\mat{J}^{\top}\mat{J}+\lambda\mat{D}_{\rm lm}$ is the
damped curvature of Eq.~\eqref{eq:lm} once data weights and any fixed
quadratic priors are folded in, and $g$ is the gradient.  Grouping the
columns of $\mat{J}$ by output--segment block realizes the exact split
concretely,
\begin{equation}
\mat{A}_{\lambda}
=
\mat{B}_{\lambda}
+
\mat{E},
\label{eq:spla_A_B_E}
\end{equation}
where the block-diagonal base $\mat{B}_{\lambda}$ collects the
within-segment curvature (one small dense factorization per block, cheap
to form and apply) and the cross-segment coupling $\mat{E}$ carries the
soft modes.  SPLA collects the troublesome directions it will treat
exactly as the columns of a full-rank matrix
$\mat{Z}\in\reals^{n\times m}$ with $m\ll n$, its \emph{coarse space}
(with $m=0$ allowed).  In practice these columns are closed-form
hinge-collision modes, directions recycled from earlier linearizations
and accepted steps, or sketched spectral modes, as described below.  The
span of $\mat{Z}$ is then inverted exactly through the standard balancing
(deflation) correction~\citep{NabbenVuik2006} built on this block base,
and the complementary directions are left to PCG.  Every represented
direction becomes an eigenvalue-one eigenvector of the preconditioned
operator, so the iteration count is controlled by the coupling left
outside the span of $\mat{Z}$ rather than by the raw model size.

The choice of $\mat{Z}$ organizes the solver options into a
ladder of progressively increasing cost,
\begin{align}
\mat{Z}=\varnothing
&\quad\Longrightarrow\quad
\text{plain block-preconditioned PCG},
\nonumber\\
\mat{Z}\ \text{geometric}
&\quad\Longrightarrow\quad
\text{the SPLA two-level preconditioner},
\nonumber\\
\mat{Z}\ \text{sketched}
&\quad\Longrightarrow\quad
\text{Nystr\"om preconditioning},
\nonumber\\
\mat{Z}=\mat{E}_{\mathcal{H}}
&\quad\Longrightarrow\quad
\text{exact elimination of a hard set},
\nonumber\\
\mat{Z}=\mat{I}
&\quad\Longrightarrow\quad
\text{a direct solve},
\label{eq:spla_special_cases}
\end{align}
ranging from the cheap block preconditioner to a full direct
factorization as $\mat{Z}$ grows.  (Here $\mat{E}_{\mathcal{H}}$ denotes
the coordinate basis of a chosen ``hard'' parameter set, for which the
correction reduces to exact Schur elimination, the exact
block elimination of those parameters from the linear system.)

When using SPLA, the level that carries the solves on the AI Feynman
fits of \S\ref{sec:spla_performance} is ``$\mat{Z}$ sketched'', which is an
adaptive randomized Nystr\"om coarse
space~\citep{FrangellaTroppUdell2023} assembled from a handful of
randomized probes of the damping-whitened operator
$\widehat{\mat{A}}_{\lambda}$.  The sketch rank is self-tuned against the
damping, with no user input, so that it tracks the operator's effective
dimension $d_{\rm eff}(\lambda)$, a soft count of the curvature
eigenvalues above the damping scale (on fitted segmented models
$d_{\rm eff}$ is far smaller than $n$).  Each
sketch product is an analytic Jacobian--vector followed by a
vector--Jacobian product (\S\ref{sec:derivatives}), far cheaper than
the autograd matrix--vector products that dominate randomized sketching
in the generic setting, and cheap enough that the level runs
inside the LM loop rather than as a one-off setup cost.  The
spectral level is also structure-agnostic.  It touches the model only
through those products, so the same construction preconditions the
Gauss--Newton system of every model family of \S\ref{sec:segmented},
including those with no hinge geometry at all.  
The segment-geometry rungs of the ladder form the
escalation route for harder cases.  Collision modes for coincident
and opposed hinge pairs are written down in closed form from the hinge
parameters, with no eigensolve, and they detect exact parametrization
redundancies that no operator sketch can see.  

Every level is safeguarded.  A step returned as a certified
linear solve must pass a freshly evaluated residual test,
$\|b-\mat{A}_{\lambda}h\|\le\mathrm{atol}+\mathrm{rtol}\,\|b\|$, while an
optimistic proposal for the nonlinear update is kept only if it decreases
the true objective, $\phi(\theta+h)<\phi(\theta)$, so an inexact step is 
never mislabeled as a certified solve.

The exact-subspace condition-number theorem that governs the framework,
the constructions of the collision modes and of the Nystr\"om level, the
proposal bank and its screening, and the supporting spectral envelopes
and diagnostics are all developed in the SPLA theory
paper.

\section{Predictive Levenberg--Marquardt}
\label{sec:predictive}

The Levenberg--Marquardt algorithm~\citep{Nocedal2006} alternates
between solving the normal equations~\eqref{eq:lm} and adjusting the
damping parameter $\lambda$.  Each LM step requires either a matrix
factorization ($\mathcal{O}(n^3)$ for direct solve) or PCG iterations, both of which
can be expensive.  We introduce several \emph{predictive mechanisms} that
anticipate good parameter updates without these costly operations, each
effective in a particular phase of the solve.  We first fix the
coordinate system, the acceptance logic, and the regularized objective
the optimizer actually minimizes
(\S\ref{sec:scaling}--\S\ref{sec:evidence}), then present the predictive
mechanisms themselves and the phase of the solve in which each earns its
keep.
The optimizer also supports residual-module assembly, blockwise union
solves (joint solves over unions of parameter blocks), and a
segment-cluster restricted additive Schwarz
(RAS)~\citep{CaiSarkis1999} variant for very large segmented models, as
well as explicit measurement-uncertainty handling in both inputs and
outputs, presented with the adaptor ecosystem in \S\ref{sec:adaptors}.
These capabilities are orthogonal to the predictive-LM idea developed in
this section.

\subsection{Solved Geometry and Parameter Scaling}
\label{sec:scaling}

A recurring practical problem in LM for neural networks is ill-conditioning
from heterogeneous parameter scales (e.g.\ amplitudes $a$ vs.\ kernels $K$).
NestyNet therefore solves the LM system in a \emph{scaled coordinate system}
(``solved geometry'') to improve conditioning and make the trust-region
logic more scale-invariant.

Let $\mat{H} \approx \mat{J}^\top \mat{W}\mat{J}$ denote the Gauss--Newton
matrix (or a block/union thereof), with $\mat{W}$ the diagonal data-weight
matrix (the identity when no noise model or robust reweighting is
supplied), and let $g := \mat{J}^\top \mat{W} r$
denote the gradient of the squared-residual loss (we use $g_{os}$ with
subscripts elsewhere for the sigmoid gate factor; bare $g$ throughout this
section refers to the gradient).  We define a diagonal scaling
\begin{equation}
\mat{S} = \diag(s_1,\ldots,s_P), \qquad s_i \approx \frac{1}{\sqrt{\mat{H}_{ii}}},
\label{eq:scaling_def}
\end{equation}
with robust quantile clipping (5th--95th percentile) to prevent extreme
$s_i$ from dominating.  We then solve in
primed coordinates $\theta' = \mat{S}^{-1}\theta$ and $h = \mat{S} h'$:
\begin{equation}
\left(\mat{H}' + \lambda \mat{D}'_{\rm lm}\right) h' = -g', \qquad
\mat{H}' = \mat{S} \mat{H} \mat{S},\quad g' = \mat{S} g.
\label{eq:scaled_system}
\end{equation}
This scaling produces two kinds of column that transform differently.
Curvature-factor columns $\mat{U}$ (those of the direct
predictor below), which enter the preconditioner as
$\mat{U}\mat{U}^\top$, map covariantly, $\mat{U}\leftarrow\mat{S}\mat{U}$, since
$\mat{S}(\mat{U}\mat{U}^\top)\mat{S}=(\mat{S}\mat{U})(\mat{S}\mat{U})^\top$, which keeps
the Woodbury formulas consistent.  Coarse-space and manifold bases
$\mat{Z}$ (the SPLA coarse space of \S\ref{sec:spla} and the
local-manifold basis below),
which enter as steps $h=\mat{Z}\alpha$, instead map contravariantly,
$\mat{Z}\leftarrow\mat{S}^{-1}\mat{Z}$, because $\operatorname{range}(\mat{Z})$
is a space of parameter displacements. Only then are the projected reduced-space
and balancing systems invariant under the scaling.  Consequently a hinge column
reused as both a curvature factor and a search direction must be transformed
according to its role in each solver.

The implementation distinguishes between the damping geometry (the choice of
$\mat{D}'_{\rm lm}$) and the acceptance and $\lambda$-update rule.
The default and recommended choice is Marquardt damping.
In the scaled system~\eqref{eq:scaled_system}, the damping diagonal is
\[
\mat{D}'_{\rm lm} = \mat{S}\,\diag(\mat{H})\,\mat{S}.
\]
Because $\mat{S} \approx \diag(\mat{H})^{-1/2}$ up to robust clipping, this choice is
close to the identity in solved coordinates while remaining adapted to the
heterogeneous local curvature in the original parameters.

\subsection{Acceptance Test, Predicted Reduction, and $\lambda$ Updates}
\label{sec:accept}

After computing a candidate step $h$, the optimizer evaluates the actual
objective decrease $\Delta\phi = \phi(\theta) - \phi(\theta+h)$ and compares
it with the quadratic model prediction.  In the unweighted case,
\begin{equation}
\widehat{\Delta\phi} = - g^\top h - \tfrac12 h^\top (\mat{J}^\top\mat{J}) h,
\end{equation}
with the weighted analogue obtained by replacing $\mat{J}^\top\mat{J}$ by
$\mat{J}^\top \mat{W}\mat{J}$.

\paragraph{Predicted reduction without explicit $\mat{J}^\top\mat{J}$.}
When solving in scaled coordinates and/or when only an implicit operator is
available, the code uses an equivalent identity derived from the LM normal
equations.  If $(\mat{H}' + \lambda \mat{D}')h'=-g'$ in solve coordinates, then
\begin{equation}
\widehat{\Delta\phi}
= -{g'}^\top h' - \tfrac12 {h'}^\top \mat{H}' h'
= \tfrac12 \left( \lambda\, {h'}^\top \mat{D}' h' - {g'}^\top h' \right) .
\label{eq:pred_red_scaled}
\end{equation}
This expression avoids explicitly forming $\mat{H}'$ and is particularly useful in the overdamped regime.

\paragraph{Trust ratio and damping schedule.}
We define the trust ratio $\rho = \Delta\phi/\widehat{\Delta\phi}$ and
record it as a diagnostic throughout, but in the controller used here $\rho$
does not determine step acceptance. A full LM step is accepted whenever it
decreases the true objective, with an Armijo backtracking line search~\citep{Armijo1966} along
the LM direction otherwise, and $\lambda$ is updated conservatively after
accept or reject.  On the stiff multiscale problems considered, $\rho$ is
too noisy to drive acceptance reliably.

\subsection{Evidence-Based Spatial Regularization}
\label{sec:evidence}

A surrogate can match the data at the noise level
and still be wrong in a way that matters. Its residuals can organize into
spatially coherent patterns that are 
imperceptible point by point.  It is this organized part of the residual,
and the excess curvature the fit spends producing it, that generally contaminates
derivatives.  Our aim here is to combat such structured misfitting by regularizing over
\emph{patches}.  We project the standardized residuals in each small
neighborhood of input space onto the patterns a systematic misfit would
leave there (a local offset, tilt, or quadratic trend) and penalize
those projections, while structure consistent with white noise is left
alone.  The penalty strengths, and the per-segment prior strengths that
accompany them, are not tuned by hand.  They are set by maximizing the
Bayesian evidence, so the data decide how much regularization they
support.

For notational simplicity we show the scalar-output case, but vector-valued
outputs are easily handled by stacking output components into the residual vector.
(``Evidence'' throughout this subsection refers to the Bayesian Laplace
marginal likelihood of Eq.~\eqref{eq:eb_laplace} below, not to observational
data).
Let $r(\theta)\in\reals^{N}$ denote the data residual vector, and let
$\Sigma=\diag(\sigma_1^2,\ldots,\sigma_N^2)$ denote known observational
variances, with $\Sigma=\mat I$ when no heteroscedastic noise model is
supplied. We define the standardized residual (within this subsection $z$
denotes this residual, not the segment pre-activation $z_{os}$ of
Eq.~\eqref{eq:zos_def}) as
\begin{equation}
z(\theta):=\Sigma^{-1/2}r(\theta) \, .
\label{eq:eb_standardized_residual}
\end{equation}
The construction is two-level: an outer evidence loop, described below,
sets the strengths $(\eta,\tilde\alpha)$, and for fixed strengths the
inner predictive LM step minimizes the augmented least-squares
objective
\begin{align}
\Phi(\theta;\eta,\tilde\alpha)
=&\frac12\|z(\theta)\|_2^2 \nonumber\\
+&\frac12\sum_{f\in\mathcal F}\eta_f\|\mat A_f z(\theta)\|_2^2 \nonumber\\
+&\frac12\sum_{(\ell,s)\in\mathcal B}
\tilde\alpha_s^{(\ell)}(t)
\bigl(\theta_s^{(\ell)}-\mu_s^{(\ell)}\bigr)^\top
\mat R_s^{(\ell)}
\bigl(\theta_s^{(\ell)}-\mu_s^{(\ell)}\bigr),
\label{eq:eb_inner_objective}
\end{align}
where $\mathcal F=\{\mathrm{mean},\mathrm{slope},\mathrm{quad}\}$ indexes
patch-families, $\eta_f\ge 0$ are patch strengths, $\mathcal B$ denotes the
active stage-local prior blocks, and $t$ is the number of accepted LM
steps.  For a single segmented model $\mathcal B=\{(0,s)\}_{s=1}^{S}$,
whereas for the stacked dual model $\mathcal B=\{(0,s),(1,s)\}_{s=1}^{S}$.

\paragraph{Patch-term construction.}
The patch term in Eq.~\eqref{eq:eb_inner_objective} penalizes residual
components that look locally like a constant offset, a slope, or a
quadratic trend, built from fixed local neighborhoods in input space.  For
each sample $x_i$ we choose a
$k$-nearest-neighbor neighborhood $\mathcal N_i$ and form a local polynomial
design matrix $\mat \Phi_i$ from a
constant column, optional first-order columns, and optional quadratic
columns. Writing a thin QR factorization as
\begin{equation}
\mat \Phi_i=\mat U_i\mat T_i \, ,
\end{equation}
we partition the orthonormal columns of $\mat U_i$ into family blocks
$\mat U_i^{(\mathrm{mean})}$, $\mat U_i^{(\mathrm{slope})}$, and
$\mat U_i^{(\mathrm{quad})}$. If $z_i$ denotes the standardized residual
restricted to the neighborhood $\mathcal N_i$, then the contribution from
family $f$ at this sample is
\begin{equation}
\bigl\|\mat U_i^{(f)\top} z_i\bigr\|_2^2 \, .
\label{eq:eb_local_patch_penalty}
\end{equation}
Stacking these local projections over all samples $x_i$, with an optional
coverage normalization so that heavily reused samples do not dominate,
yields a sparse global operator $\mat A_f$.  Equivalently, the data
precision is replaced by
\begin{equation}
\mat Q:=\mat I+\sum_{f\in\mathcal F}\eta_f\,\mat A_f^\top\mat A_f \, ,
\label{eq:eb_Q_precision}
\end{equation}
so that the first two terms of Eq.~\eqref{eq:eb_inner_objective} combine
into $\tfrac12 z^\top\mat Q z$.  The patch
operators are built once from the sampled input geometry and then held fixed
during optimization.

\paragraph{Gauge-aware stage priors.}
The third term in Eq.~\eqref{eq:eb_inner_objective} is a Gaussian shrinkage
toward stage-local anchors, used to keep parameters near the values set by the canonical
initialization without forcing cross-stage coupling.  The shrinkage is
stage-local rather than fused across layers.  The
anchor $\mu_s^{(\ell)}$ is a declared choice. The default is the deterministic
seed initialization, a pre-data function of topology and seed alone, and when
canonical initialization has been run the anchor may instead be captured from
it, in which case it inherits that construction's dependence on the training
data and must be declared as part of the conditioning information.  The
precision template
$\mat R_s^{(\ell)}=\diag((\rho_{sj}^{(\ell)})^{-2})$ is built from
stage-local family scales:
\begin{equation}
(\rho_{sj}^{(\ell)})^{-2}
=
\left[
(\tau_{\rm rel}\,\mathrm{RMS}_{g(j,\ell)}^{(\ell)})^2+\tau_{\rm abs}^2
\right]^{-1},
\label{eq:eb_stage_family_precision}
\end{equation}
where $g(j,\ell)$ denotes the parameter family (amplitude, bias, or kernel)
containing component $j$ in stage $\ell$.  The
corresponding prior density is
\begin{align}
p(\theta\mid\alpha,\mu,\mat R)&\propto
\prod_{(\ell,s)\in\mathcal B}
\exp\!\Bigl[
-\tfrac12\tilde\alpha_s^{(\ell)}(t)\,
Q_s^{(\ell)}
\Bigr],\nonumber\\
Q_s^{(\ell)} &\equiv
\bigl(\theta_s^{(\ell)}-\mu_s^{(\ell)}\bigr)^\top
\mat R_s^{(\ell)}
\bigl(\theta_s^{(\ell)}-\mu_s^{(\ell)}\bigr) \, .
\label{eq:eb_segment_prior}
\end{align}
For a dual-layer model the first stage initially serves partly as an exact
feature lift and only gradually becomes a fully data-driven nonlinear
representation.  We therefore use stage-aware effective shrinkages
\begin{equation}
\tilde\alpha_s^{(\ell)}(t)=\kappa_s^{(\ell)}(t)\,\alpha_s^{(\ell)},
\label{eq:eb_stage_gated_alpha}
\end{equation}
with $\kappa_s^{(1)}(t)=1$ from the outset for the outer stage, while the
inner-stage factor is delayed for newly canonicalized segments and then
turned on with a reduced scale $0<\kappa_s^{(0)}\le 1$.

\paragraph{Laplace evidence and outer loop.}
To score the regularization strengths $\eta_f$ and $\alpha_s^{(\ell)}$
without trial-and-error, we use the Laplace approximation to the marginal
likelihood (``Bayesian evidence''), following the evidence framework
of~\citet{MacKay1992}.
Let $\mu$ collect all stage-local anchors and define the block-diagonal
prior precision
\begin{equation}
\mat P(t):=\mathrm{blockdiag}\!\Bigl(
\tilde\alpha_s^{(\ell)}(t)\mat R_s^{(\ell)}
\Bigr)_{(\ell,s)\in\mathcal B}.
\label{eq:eb_P_precision}
\end{equation}
If $\mat J_z=\partial z/\partial\theta$ denotes the Jacobian of the
standardized residual, then the Laplace/Gauss--Newton Hessian of the inner
problem is approximated by
\begin{equation}
\mat H \approx \mat J_z^\top \mat Q \mat J_z + \mat P(t).
\label{eq:eb_H_precision}
\end{equation}
Up to additive constants independent of $\theta$, the corresponding Laplace
log evidence is
\begin{align}
\log p(y)
\approx
-&\frac12 z^\top\mat Q z
-\frac12(\theta-\mu)^\top\mat P(t)(\theta-\mu) \nonumber\\
&+\frac12\log|\mat Q|
+\frac12\log|\mat P(t)|
-\frac12\log|\mat H|.
\label{eq:eb_laplace}
\end{align}
The expression assumes that $\mat Q$ and $\mat P(t)$ are positive definite.
When warm-up disables some shrinkages and $\mat P(t)$ is singular,
$\log|\mat P(t)|$ is taken over the active regularized subspace (a
pseudo-determinant, omitting the corresponding flat-prior constants), and
such evidences are compared only at fixed active dimension. Alternatively,
a small positive floor on the shrinkages keeps all priors proper.
Operationally this still yields a clean separation of roles: the inner loop
remains an LM solve of the augmented least-squares objective
\eqref{eq:eb_inner_objective}, while the outer evidence loop scores the
regularization strengths through the Laplace approximation.

\subsection{Quadratic Path Extrapolation}

The sequence of accepted parameter vectors
$\theta^{(0)}, \theta^{(1)}, \ldots, \theta^{(k)}$ traces a smooth path
through parameter space.  Late in the optimization process, this path often follows a
predictable, low-curvature trajectory toward the minimum, and we exploit
this regularity by fitting a quadratic model to the parameter history:
\begin{equation}
\theta(t) = \vect{a} + \vect{b}t + \vect{c}t^2
\label{eq:quadratic}
\end{equation}
where $t \in \{0, 1, \ldots, k\}$ indexes the accepted iterates.  The
coefficients $(\vect{a}, \vect{b}, \vect{c})$ are determined by least-squares
on the last $N_{\rm history}$ iterates (typically $N_{\rm history} = 10$).

At each step we extrapolate forward by $\Delta t > 0$ steps,
\begin{equation}
\theta_{\rm pred} = \theta(k + \Delta t) = \vect{a} + \vect{b}(k+\Delta t) + \vect{c}(k+\Delta t)^2 \,,
\end{equation}
evaluate the loss $\phi(\theta_{\rm pred})$, and accept the extrapolated
point whenever $\phi(\theta_{\rm pred}) < \phi(\theta^{(k)})$.
The jump size $\Delta t$ is adapted based on success. On acceptance we grow
it by 10\% ($\Delta t \leftarrow 1.1 \cdot \Delta t$) and on rejection we
shrink it by 10\% ($\Delta t \leftarrow 0.9 \cdot \Delta t$).
Each extrapolation attempt costs only a forward pass, so even moderate
acceptance rates substantially reduce the expensive matrix operations.  We
quantify the contribution in \S\ref{sec:benchmarks}.

\subsection{Cheap Step Predictors: Woodbury Fast Path and Local Manifold Solve}
\label{sec:predictors}

Before committing to an expensive factorization or PCG solve, the
optimizer attempts two cheap predictors, each judged by the same
safeguards as every other proposal.  The \emph{direct predictor}
approximates the damped system by a diagonal-plus-low-rank model whose
curvature-factor columns $\mat{U}$ carry the dominant cross-segment
curvature that the diagonal omits.  The Woodbury identity then collapses
the solve to a small system of the retained rank, at cost linear in $n$.
The candidate step is accepted only if it passes a damping-aware
residual test with threshold $\tau(\lambda)$. At large $\lambda$ the
diagonal-plus-low-rank model is nearly exact and the expensive solve is
skipped entirely, while as $\lambda\to0$ the threshold tightens and the
solver falls back to full PCG.  This fast path is off by default and is
enabled per problem.  The \emph{local manifold solve} instead projects
the LM system onto a small orthonormalized basis $\mat{Z}$ assembled
from the normalized gradient, the SPLA coarse directions of
\S\ref{sec:spla}, and recent accepted steps, and solves the reduced
system exactly, with an optional safeguarded correction from the exact
residual curvature.  The gradient, hinge columns, and recent steps
together span most of the useful step directions, so this small solve
often succeeds at negligible cost.  Both constructions, their acceptance
thresholds, and their variants are developed in the SPLA theory paper.

\subsection{L-BFGS Rescue}

The cheap predictors above are designed to bypass expensive linear algebra
when the local LM model is already informative.  A different failure mode
arises in narrow, winding valleys, where LM can spend many outer iterations
rejecting misaligned steps or making only weak progress along the valley
floor.  For such regimes the optimizer includes a limited-memory
BFGS~\citep{Nocedal1980,LiuNocedal1989} rescue.  When recent LM history
indicates local stagnation, curvature pairs (successive parameter and
gradient differences) harvested from accepted iterates
(i.e.\ at no additional computational cost) are used to propose a
quasi-Newton step, which is then subjected to the same objective-decrease
and safeguard tests as an LM proposal.  This provides a second-line 
escalation path when the local quadratic model
is poorly aligned with the long curved geometry of the objective.
In the experiments reported here the rescue runs with fixed default settings
throughout: it is triggered after five consecutive rejected LM steps, draws on a
history of ten curvature pairs, takes at most twenty L-BFGS iterations per
attempt, and is limited to ten such attempts per run.  

\subsection{Stage-Specialization of the Predictive Mechanisms}

The predictive mechanisms are \emph{stage-specialized}. Each is effective only in a
particular phase of the solve, so a single averaged ``speedup'' conflates regimes in
which a mechanism is decisive with regimes in which it is idle or even
counterproductive.  Quadratic path extrapolation is essentially inactive during the
early, large-step phase (the accepted iterates do not yet trace a smooth,
low-curvature path) and engages only in the small-$\lambda$ endgame, where it
fast-forwards the slow grind of diminishing improvements toward the optimum.  Post-step linear refinement is its complement in time.  Eliminating the
linearly entering parameters at each step (a variable-projection-style
move~\citep{Golub1973,Golub2003}) accelerates the early approach, but the
same eagerness can perturb a clean superlinear Gauss--Newton endgame, so
its value peaks in the approach phase and can vanish, or reverse, near
convergence.  Section~\ref{sec:predictor_ablation} illustrates the
stage-split on a representative problem.  Every proposal is safeguarded
as in \S\ref{sec:accept}, and each mechanism tracks a
rolling acceptance rate and is disarmed when that rate falls, so its cost
outside its regime stays negligible.

The optimizer also supports trust-region dogleg fallbacks and geodesic
acceleration~\citep{Transtrum2011,Transtrum2012}, the latter improving
robustness in curved valleys rather than bypassing steps.  The
L-BFGS rescue is aimed at rare stagnation regimes rather than
average-case throughput.

\section{Modular Ecosystem for Scientific Optimization}
\label{sec:adaptors}

Real-world physics problems often require composing multiple model
components: outer transforms, PDE constraints, implicit solves, and linear
projections.  NestyNet provides a composable framework through a common
interface that exposes residuals, matrix-free derivative operators
(Jacobians and Hessians applied through their vector products, never
instantiated), parameter blocks, and optional low-rank preconditioner inputs.  This is the
mechanism that lets the same optimizer handle direct fitting, PDE losses,
variable-projection templates~\citep{Golub1973}, and implicit Hamilton--Jacobi constructions
without giving up analytic derivatives.  An autograd adaptor completes
the interface, wrapping any PyTorch module as a provider with
derivatives supplied by automatic differentiation, so models outside
the segmented family plug into the same optimizer.

The same machinery also extends to complex-valued segmented models.  In
that setting the parameters, activations, and Jacobians are complex, but the
optimization objective remains a real least-squares functional built from
residual magnitudes.  Accordingly, the Gauss--Newton and LM systems use the
Hermitian adjoint $\mat{J}^\dagger$ in place of the ordinary transpose (the
derivatives taken in the standard Wirtinger sense for the real objective
$\tfrac{1}{2}\|r\|^{2}$, i.e.\ treating the complex parameters and their
conjugates as independent variables),
allowing the same analytic-derivative framework to optimize complex-valued
surrogates without reducing them to real-imaginary splits.

This ``adaptor'' architecture is part of the contribution of this first paper, and
is what makes the segmented surrogate and predictive LM optimizer usable as a
general scientific least-squares substrate.  Here we use and validate only a
small subset of it (the single and dual stacked adaptors, and the
PDE segmented adaptor with periodic and Neumann boundary conditions). 
The released framework currently includes over 45 adaptor classes.
Composition adaptors provide single, dual, and arbitrary-depth sequential
stacks, ResNet-like (skip-connection) blocks, concatenation, and
parameter-subspace masks.
Input and output transforms cover log, asinh, logit, Box--Cox, and
Yeo--Johnson maps, output links, coordinate maps with analytic chain-rule
corrections, and learnable outer functions for singularities, while
residual transforms supply both the patch residuals used by the evidence-based
whitening and the curvature-supervised Sobolev-gradient residuals.  The
physics families span PDE residuals and boundary conditions
(time-dependent PDEs, Fokker--Planck, Euler--Lagrange, periodic, and
Neumann), dynamics and integration (multiple shooting, leapfrog,
Lagrangian-step variants, symplectic shear, and orbital adjoints),
Hamilton--Jacobi providers (implicit HJ, pair-actions, angle-linearity,
learnable Hamiltonians, and action-angle), variable projection, and a
SINDy (sparse dynamics identification) library~\citep{Brunton2016}.

Measurement uncertainties are handled explicitly on both sides of the
regression.  On the observation side, errors enter as per-point variances
or whitening operators, so correlated errors among the outputs of an
observation are accommodated, and robust reweighting covers non-Gaussian
tails.  The errors-in-variables adaptors support input uncertainties at
two levels, from effective-variance weighting built on the analytic input
Jacobian (\S\ref{sec:derivatives}) to full orthogonal-distance de-biasing,
in which per-point latent input corrections are fitted jointly with the
parameters.  Measurement errors in $x$ and $y$, including their
correlations, are thus combined in one solver.  These adaptors carry
measurement uncertainty into the fit.  The uncertainty information flowing
out of a converged fit is the subject of Paper~II.

The de-biasing above is one instance of a general mechanism.  The
library eliminates blocks of variables from the Gauss--Newton step in
two statistically distinct ways that share the same Schur complement.
Solver eliminations, such as variable projection and the hard-set rung
of the \S\ref{sec:spla} ladder, remove parameters of interest to
condition the solve, and restore them afterwards.  Statistical
marginalizations remove nuisances that never rejoin the parameter
vector, per-point latent input corrections as above or calibration
offsets shared by groups of observations, and the reduced curvature is
then the correct curvature to report.

The same adaptor interface also applies constraints.  Masking adaptors 
can freeze user-selected
parameters, segments, or outputs of any provider, and the optimizer can
enforce box, linear-equality, and inequality constraints on any parameter
subset, all without leaving the analytic-derivative path.

The framework also extends to forward-modeling of the observation process
itself.  A response-operator layer fits data observed through selection or
completeness functions, binning, masks, point-spread convolution, or
line-of-sight projection, with the operator adjoints carried by the same
analytic-derivative path.  Companion likelihood families expressed as
deviance residuals (signed square roots of each datum's log-likelihood
deficit) let the predictive LM solver optimize the exact likelihood of
Poisson counts, injection--recovery calibration, censored measurements,
and correlated Gaussian noise while remaining a least-squares method.
Because window-limited data determine only part of the reconstruction,
the layer includes the corresponding identifiability diagnostics.  This
observation-model layer will be presented in detail in a subsequent
contribution.

\section{Application to PDE Solutions}
\label{sec:applications}

Partial differential equations provide a natural stress test of the main
claim of this contribution.  One must fit not only the solution field
itself, but also its local differential structure, and the induced
optimization landscape is often stiff, multiscale, and strongly correlated.
In the standard PINN formulation~\citep{Raissi2019}, one solves
\begin{equation}
\min_\theta \|r_{\rm PDE}\|^2 + \alpha\|r_{\rm BC}\|^2,
\qquad
r_{\rm PDE} = \mathcal{L}[u_\theta] - f,
\end{equation}
where $\mathcal{L}$ is the differential operator, $f$ the source term,
$r_{\rm BC}$ the boundary-condition residual, and $\alpha$ a balance
weight.

For the purposes of the present paper, we examine whether NestyNet can supply a useful
surrogate-and-optimizer pair for this regime.  The analytic derivative
machinery of \S\ref{sec:derivatives} makes operators such as the
Laplacian available in closed form and at practical cost, while the
predictive Levenberg--Marquardt solver attacks the resulting
least-squares problem in a way that is explicitly designed for strongly
correlated parameter geometries.  In other words, the same ingredients that
make NestyNet effective on difficult scientific functions also make it a
credible substrate for PDE residual minimization.

We tested this directly against the PINNacle
benchmark~\citep{PINNacle2024}, which provides 22 standardized PDE problems
with fixed domains, boundary conditions, and evaluation protocols.  On two
of the 22 cases (PNd and HNd, PINNacle's high-dimensional Poisson and
Heat stress tests, both in 5 spatial dimensions), NestyNet surpasses the
best result reported in the benchmark.

For PNd, a single-stage 32-segment model is first warmed up with Adam
(the default first-order optimizer of deep learning;
\citealp{KingmaBa2015}) and
then polished with exact Levenberg--Marquardt.  In a three-seed protocol the
Adam-to-LM route reaches a validation error of $4.0\times 10^{-7}$, and LM
polish outperforms Adam continuation on every seed.  For HNd, a 32-segment
model with only $1\,536$ parameters reaches $1.08\times 10^{-4}$ on the
official $20\,000$-point validation set.  Averaged over 10 independent
random-geometry draws of the same size, the error is $1.09\times 10^{-4}$
(range $1.06$--$1.13\times 10^{-4}$), which is slightly below the
benchmark's best reported result of $1.19\times 10^{-4}$.

We regard these as two genuine transfers of the segmented-model-plus-LM
recipe to the PDE setting, while noting that many of the remaining $20$ PINNacle
cases will likely require problem-specific design beyond the scope of this
paper.  Competitive physics-only solutions require careful problem-specific
collocation, boundary enforcement, and loss balancing that we have not
attempted here.  We can nevertheless check whether the segmented
representation itself is sufficiently expressive, by fitting the problems
under supervision with five frozen data seeds per case and canonical
initialization throughout.  Where analytic or independently constructed
solution values exist we fit those, and otherwise we fit predeclared
held-out subsets of the supplied reference tables.
Nineteen of the twenty are better than the quoted best solution value on every seed,
and Poisson2d-C is so on four of five against its oracle (an
independently constructed reference solution).
These supervised fits are of course not PINN solutions, and the strength
of the evidence varies with the per-case protocol, but they show that
under supervision the segmented design reaches the error scale the
benchmark requires.  The open challenge is reaching that solution from the physics
alone.  Full per-case protocols and results accompany the released
library.

\section{Numerical Experiments and Ablation Studies}
\label{sec:benchmarks}

We now present numerical experiments comparing NestyNet against baseline
methods, organized as eight complementary studies, including ablations, so that the reader may
judge each architectural and algorithmic choice independently.
The test bed throughout is the 120-equation AI Feynman benchmark as
standardized in SRBench~\citep{Udrescu2020,LaCava2021}, which spans
scalar physics equations from mechanics and electromagnetism to quantum
mechanics and thermodynamics.
Unless noted otherwise, the optimizer settings are
held at the defaults of \S\ref{sec:predictive} throughout.

\subsection{Same-Model Optimizer Comparison}
\label{sec:same_model}

\begin{figure}[t]
\centering
\includegraphics[width=0.98\columnwidth]{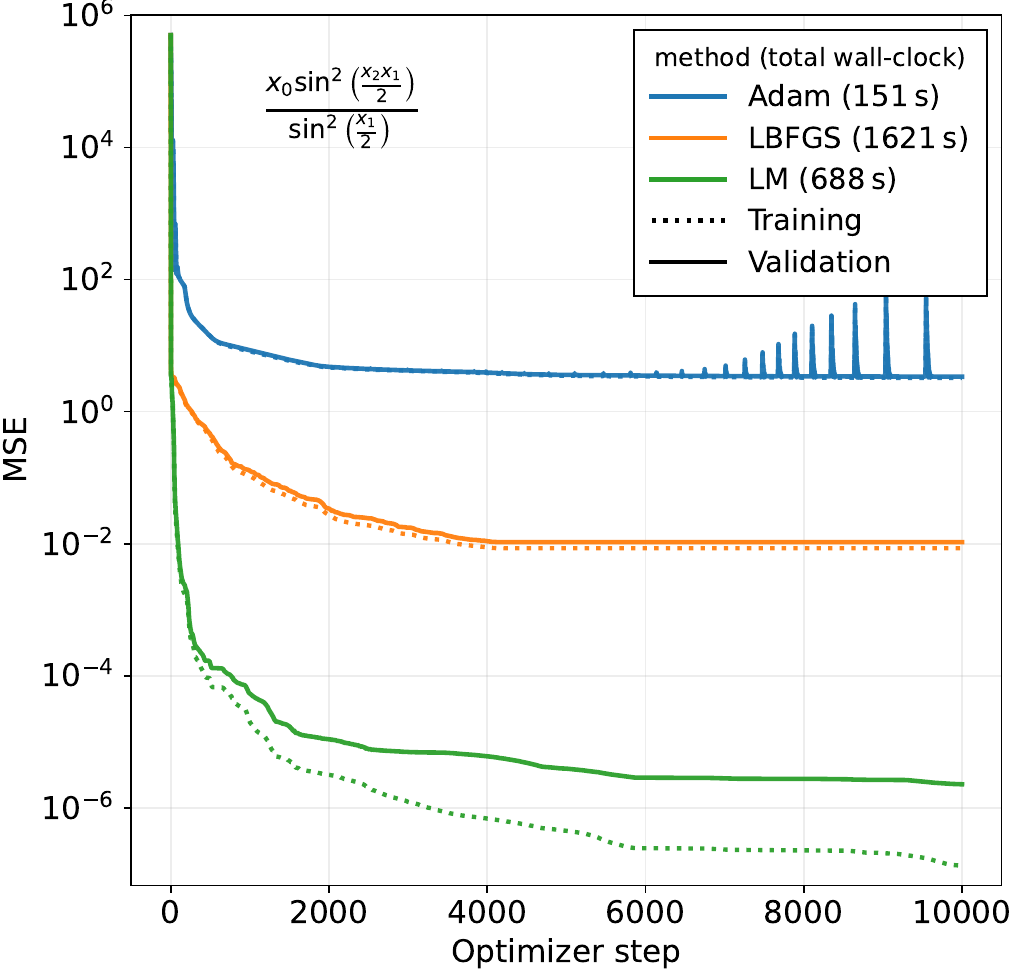}
\caption{Optimizer comparison on the same 960-parameter (dual-layer) segmented
model fit to AI Feynman \#29, from identical initialization and
with the same full-batch forward-MSE objective.  Dotted lines show training MSE
and solid lines show validation MSE, and the legend lists each method's total
optimizer wall-clock time.  Predictive LM reaches the lowest error by a wide
margin ($2.27\times 10^{-6}$ best validation MSE in $688$~s, versus $1.05\times 10^{-2}$ in $1\,621$~s
for L-BFGS and $3.38$ in $151$~s for Adam).  The difficulty is clearly not the closed form of the law but the
optimizer geometry induced by fitting a large nonlinear surrogate to it.}
\label{fig:LM_Adam_LBFGS_compared}
\end{figure}

Figure~\ref{fig:LM_Adam_LBFGS_compared}
compares Adam, L-BFGS, and predictive LM on exactly the same
960-parameter dual-layer segmented model, trained on $5\times10^3$ points from
identical initialization against the same forward mean-squared-error
(MSE) objective, evaluated full-batch (every step uses all training
points), with each method's total wall-clock time reported in the legend.  The fitted
target is the AI Feynman \#29 function
\begin{equation}
f(x_0,x_1,x_2)=\frac{x_0 \sin^2(\frac{x_2 x_1}{2})}{\sin^2(\frac{x_1}{2})},
\end{equation}
which is symbolically simple but numerically awkward because narrow curved
valleys and near-singular structure appear in the induced loss landscape.

This same-model comparison reinforces the central point of this paper.
Difficult scientific targets induce stiff, correlated optimization geometry
for which first-order training is the wrong tool.  L-BFGS improves
dramatically over Adam, as expected if curvature matters, and on simpler problems it
would also hold the practical advantage over LM, since it captures useful
curvature without forming or solving the full normal equations.  Because
the segmented surrogate exposes analytic derivatives and structured linear
algebra, however, predictive LM can use the residual geometry directly and
still be faster.

\subsection{Initialization, Segment-Prior, and Prior-Decay Ablation}
\label{sec:init_ablation}

We also evaluated an eight-configuration AI Feynman ablation crossing two model
families (single- and dual-layer) with four training recipes: plain,
canonical initialization, canonical with a persistent segment prior, and
canonical with a smoothly decayed prior (annealed between iterations 800
and $1\,000$ of $2\,000$).  Over the 88 problems retained after excluding targets
with dynamic range above 4 decades (treated in Paper~III
via an $\operatorname{asinh}$ transform), the prior-decay variant was best
in both families, reducing the geometric-mean best validation MSE by
factors of $3.90$ and $4.19$ over the plain baselines. The segment prior is
most useful as an early guide into a good basin, annealed away once the
optimizer reaches a productive region.

\subsection{Derivative Accuracy on AI Feynman}
\label{sec:derivative_accuracy}

The benchmark's scalar targets are its restriction, not the method's. The
multi-output path is exercised by the coupled vector-valued fits of
\S\ref{sec:applications} and by the vector-field systems of
Paper~IV.

All derivative comparisons use $1.5\times10^4$ training points and
$5\times10^3$ validation points drawn from the same SRBench-standardized 
domains and shared across NestyNet, Adam MLP, and L-BFGS-refined MLP fits. 
The MLP baseline is a fixed two-hidden-layer tanh-activated network with 64 units 
per layer, trained in double precision.  L-BFGS denotes full-batch refinement 
initialized from the corresponding Adam checkpoint, with no change in architecture, 
data split, precision, or preprocessing.  The NestyNet fits use dual-layer models
with 20 segments.  They are produced by the derivative-suite
runner released with this paper, using the same canonical initialization and
fitting contract as Paper~III (but before any separability analysis or symbolic rewriting).  All 120
entries are selected from this single declared suite; we do not
substitute equation-specific reference refits.  Targets selected by the
automatic loss-conditioning policy use an $\operatorname{asinh}$ weighting of
the $y$ residuals. For those cases the Adam and L-BFGS baselines are also
trained in the matching transformed coordinate.  The complete checkpoints and
row-level provenance are supplied with the code release.

\paragraph{Choice of baselines.}
We compare against neural surrogates rather than classical interpolants
because the comparison is scoped to the question this framework addresses.
Smoothing splines, radial basis functions, Gaussian processes, and
polynomial or rational bases are powerful tools for low-dimensional, scalar,
smooth interpolation, and several furnish analytic derivatives as well.  On a
smooth three-dimensional case such as AI Feynman problem~\#29, a well-tuned Gaussian
process (GP) and the dual-layer segmented surrogate, each fitted to the same
$5\times10^3$ points, reach essentially the same accuracy: relative errors
of $0.03\%$, $0.25\%$, and $1.2\%$ (GP) versus $0.04\%$, $0.31\%$, and
$1.1\%$ (NestyNet) on values, gradients, and second derivatives, so the
segmented surrogate matches the strongest classical interpolant even in the
regime that most favors it.  The distinction lies elsewhere.  The Gaussian
process is fitted by a dense kernel solve that is $\mathcal{O}(N^2)$ in
memory and $\mathcal{O}(N^3)$ in time (a scaling confirmed
empirically on this problem). Its kernel matrix alone reaches
$\sim\!80$\,GB at $N=10^5$, so it cannot reach the large-sample regime
in which the segmented surrogate's cost
grows only linearly in $N$.  More fundamentally, kernel and spline
interpolants are not designed to be \emph{composed} (stacked, embedded in
PDE residuals or implicit solves, tied across vector-valued outputs, and
optimized end-to-end by a second-order method), which is the substrate role
developed here and exploited in Papers~III and~IV.  The neural-surrogate
comparison in this section therefore isolates the property at issue: given a
reusable, composable surrogate, which architecture yields trustworthy
differential operators?

\begin{figure*}[t]
\centering
\includegraphics[width=0.96\textwidth]{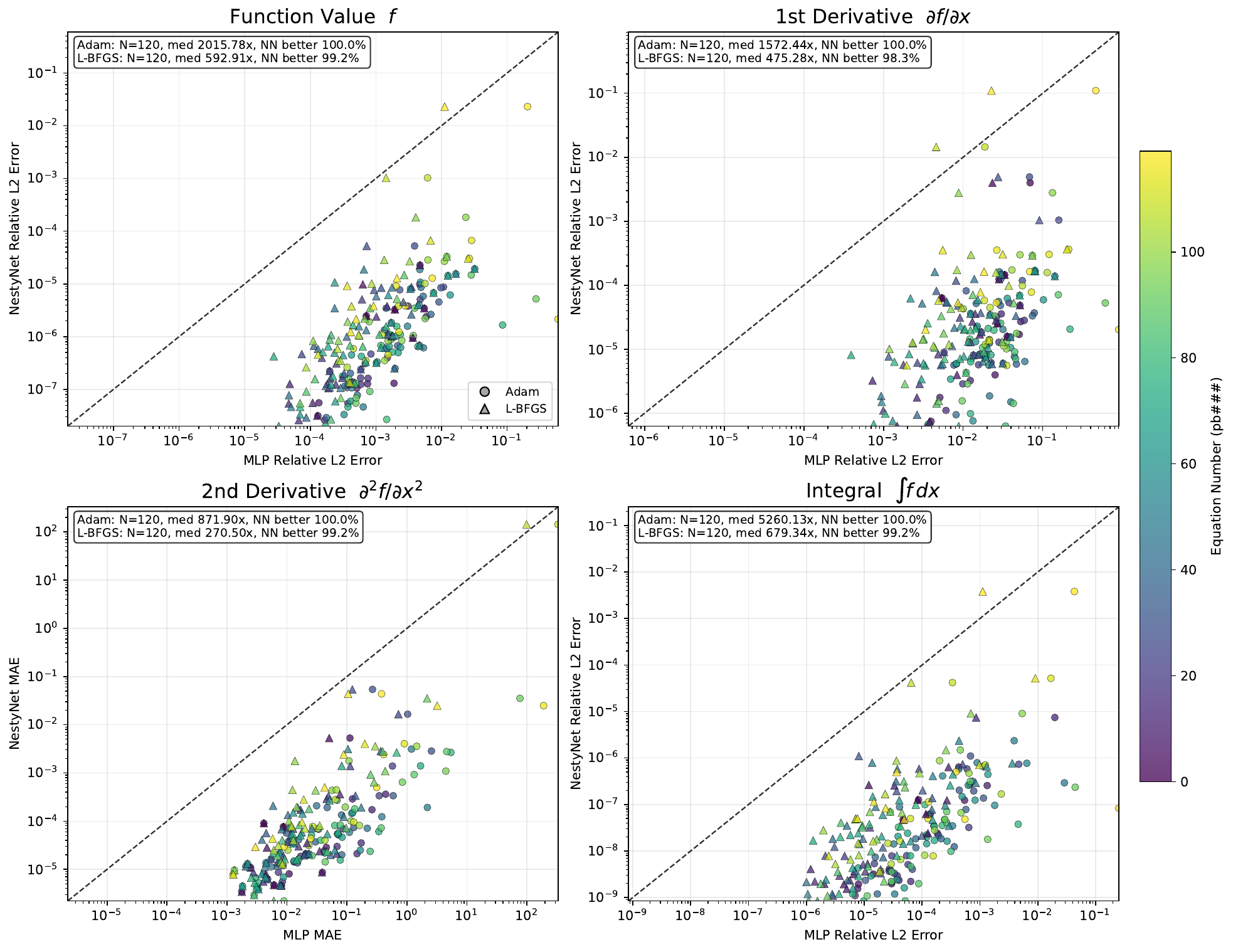}
\caption{Direct MLP-vs-NestyNet error comparison across the benchmark suite.
Circles show Adam-trained MLPs, while triangles show
full-batch L-BFGS runs initialized from the corresponding Adam solution.  The
dashed line marks parity, so points
below the line favor NestyNet.  For the derivative and integral panels, each
plotted point is the per-equation median over input axes.  The panels show relative $L^2$ errors for values, first derivatives, and
integrals, and MAE for second derivatives (see text).  The median MLP/NestyNet improvement
factors are $2\,016\times$ (values), $1\,572\times$ (first derivatives),
$872\times$ (second derivatives), and $5\,260\times$ (integrals) for Adam, and, in
the same order, $593\times$, $475\times$, $271\times$, and $679\times$ after L-BFGS
refinement.  NestyNet performs better than Adam on $100\%$ of
cases, and better than L-BFGS on $98.3$--$99.2\%$ of cases, depending on the
panel.
Color encodes the benchmark equation index.}
\label{fig:adam_vs_nestynet_scatter}
\end{figure*}

\paragraph{Standard derivatives.}
Figure~\ref{fig:adam_vs_nestynet_scatter}
shows a direct head-to-head scatter of relative $L^2$ errors\footnote{Here the relative
$L^2$ error of a quantity $q$ is
$\Vert\hat q-q\Vert_{L^2}/\Vert q\Vert_{L^2}$, the root-mean-square of the
error over the held-out validation points normalized by the
root-mean-square of the target.}
in function
values, first derivatives, and integrals, together with the second-derivative
mean absolute error (MAE), across the benchmark suite.  We use MAE in the
second-derivative panel because
many benchmark directions have exactly zero or numerically tiny diagonal
curvature, for which normalized relative metrics become ill-conditioned and
can explode even when the absolute curvature error remains small.  In the
integral panel, all models are deliberately integrated with the same
trapezoid estimator, rather than with the closed-form antiderivatives of
\S\ref{sec:antiderivatives}, so that the comparison isolates fit quality
rather than integration machinery.

Figure~\ref{fig:adam_vs_nestynet_scatter} shows that the improvement is
remarkably consistent across equations.  The reason is straightforward.
The MLP baselines are trained only on function values, so their derivatives
are implicit byproducts of the fit, whereas NestyNet differentiates an
analytic surrogate whose local differential structure is built into the
representation.  Full-batch L-BFGS, initialized from the Adam solutions,
substantially improves the MLPs, but the derivative and integral errors remain
typically more than two orders of magnitude above NestyNet.

\paragraph{Operator probes.}
For scientific applications (including PINNs) it is also useful to ask whether the gain
persists for operator probes not tied to coordinate-aligned derivatives.  A
second benchmark view therefore uses four quantities closer to the objects
entering scientific residuals:
 a directional derivative
$v\!\cdot\!\nabla f$, mixed second partials $\partial_i \partial_j f$ with
$i<j$ (this panel shows $N=119$ problems because one equation has a single input and
therefore no off-diagonal partials), directional curvature $v^\top H v$, and the Laplacian $\nabla^2 f$.
Here $v$ is a fixed unit probe direction for each evaluated sample, shared
between the Adam, L-BFGS, and NestyNet comparisons.

\begin{figure*}[t]
\centering
\includegraphics[width=0.96\textwidth]{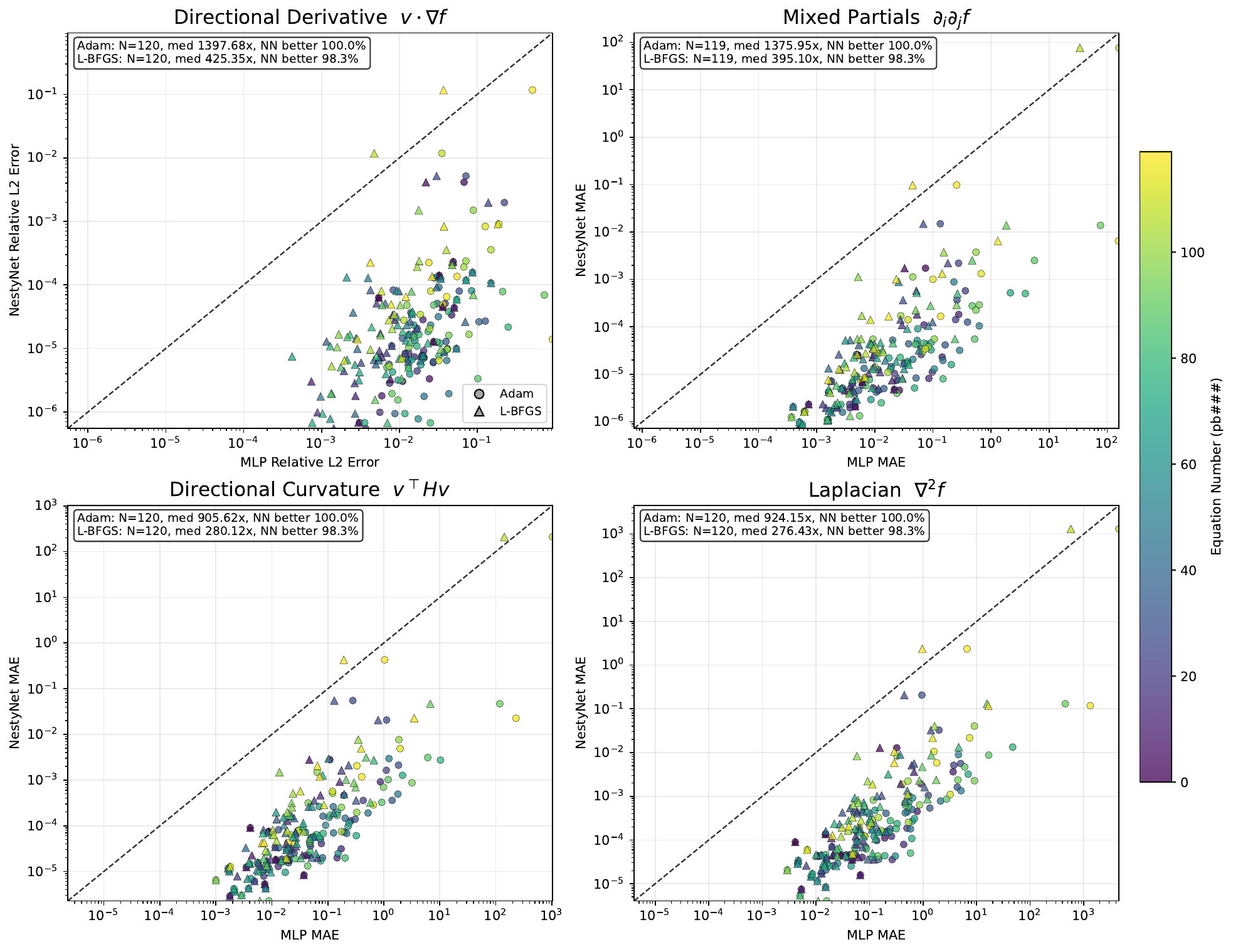}
\caption{Operator-focused MLP-vs-NestyNet comparison across the benchmark
suite.  Symbols as in Figure~\ref{fig:adam_vs_nestynet_scatter}.  The panels show the relative $L^2$ error
in a directional derivative $v\!\cdot\!\nabla f$, together with MAE for mixed
partials $\partial_i \partial_j f$ ($i<j$), directional curvature
$v^\top H v$, and the Laplacian $\nabla^2 f$.  For the mixed-partial panel,
each plotted point is the per-equation median MAE over all off-diagonal pairs
$i<j$, so higher-dimensional equations do not dominate the summary.  MAE is used for the three second-order panels for the reason given in
Figure~\ref{fig:adam_vs_nestynet_scatter}.  The median MLP/NestyNet improvement factors are
$1\,398\times$, $1\,376\times$, $906\times$, and $924\times$ for Adam, and
$425\times$, $395\times$, $280\times$, and $276\times$ after L-BFGS refinement,
respectively.  NestyNet performs better than Adam on $100\%$ of
the plotted cases in every panel, and better than L-BFGS on $98.3\%$ of cases in every
operator.  Color encodes the benchmark equation index.}
\label{fig:adam_vs_nestynet_operator_scatter}
\end{figure*}

Figure~\ref{fig:adam_vs_nestynet_operator_scatter} shows that the advantage is
not an artifact of probing only axis-aligned derivatives.  NestyNet remains
systematically better with off-diagonal curvature, contracted Hessian
probes, and Laplacians, including relative to the L-BFGS-polished MLPs.  This
is exactly the regime relevant for PDE residuals and boundary operators.

Stratified by input dimension, the median mixed-partial improvement over
the practically most relevant 2D--4D subset is $1\,566\times$ for Adam and
$481\times$ after L-BFGS refinement, versus $856\times$ and $267\times$ for
equations with five or more inputs.

\subsection{An Interpolation Law for the Derivative Gap}
\label{sec:derivative_gap}

The comparisons above document a large but remarkably regular
\emph{derivative gap}: for every baseline, the gradient improvement factors lie
between the value and curvature improvement factors.  This regularity is
the visible form of an elementary spectral inequality that every fitted model
in the suite obeys to within a factor of order unity, which (remarkably)
allows us to convert value accuracy plus curvature control into gradient accuracy.

Let $e := \hat f - f$ denote the error of a fit $\hat f$ to a target $f$,
and define the first- and second-order Sobolev (semi)norms
\begin{equation}
\Vert\nabla e\Vert_{L^2}^2=\sum_{i}\Vert\partial_i e\Vert_{L^2}^2,
\qquad
|e|_{H^2}^2=\sum_{i,j}\Vert\partial_i\partial_j e\Vert_{L^2}^2 .
\label{eq:gap_seminorms}
\end{equation}

\begin{proposition}[Gradient interpolation bound]
\label{prop:derivative_gap}
For $e\in H^{2}(\reals^{N_x})$,
\begin{equation}
\Vert\nabla e\Vert_{L^2}^{2}
\;\le\;
\Vert e\Vert_{L^2}\;|e|_{H^2},
\label{eq:gap_interpolation}
\end{equation}
with equality if and only if the power spectrum $|\hat e(k)|^{2}$ of the
error is concentrated on a single wavenumber shell $|k|=k_e$.  On
$\reals^{N_x}$ that shell has Lebesgue measure zero, so the only $H^2$ function
attaining equality is $e=0$ and the inequality is strict for every nonzero
error.  Nonzero equality is available on periodic or discrete spectra, where the
Fourier coefficients can sit on one Laplacian eigenshell, and on
$\reals^{N_x}$ it is approached as the Fourier power concentrates in a thin
annulus.  Moreover
$|e|_{H^2}=\Vert\nabla^2 e\Vert_{L^2}$: the Frobenius--Hessian seminorm
coincides with the $L^2$ norm of the Laplacian of the error.
\end{proposition}

To see this, note that by Parseval's Theorem,
$\Vert\nabla e\Vert_{L^2}^{2}=\int |k|^{2}|\hat e(k)|^{2}\,\dd k$.
Cauchy--Schwarz applied to the factors $|\hat e|$ and $|k|^{2}|\hat e|$
gives
\begin{equation}
\int |k|^{2}|\hat e|^{2}\dd k
\le
\Bigl(\int |\hat e|^{2}\dd k\Bigr)^{\!1/2}
\Bigl(\int |k|^{4}|\hat e|^{2}\dd k\Bigr)^{\!1/2},
\end{equation}
and $\sum_{ij}k_i^2 k_j^2=|k|^4$ identifies the last factor with
$|e|_{H^2}^2$, equivalently with $\Vert\nabla^2 e\Vert_{L^2}^2$ since
$\widehat{\nabla^2 e}=-|k|^2\hat e$.  Equality in Cauchy--Schwarz requires
$|k|^{2}|\hat e|\propto|\hat e|$ almost everywhere, that is, spectral
support on a single shell.

Inequality~\eqref{eq:gap_interpolation} belongs to the classical
Landau--Kolmogorov family of interpolation inequalities between
derivative norms~\citep{HardyLittlewoodPolya1952}.  What is new 
below is not the bound but how close the fitted models lie to equality.
We therefore define the tightness ratio
\begin{equation}
\rho_e := \frac{\Vert\nabla e\Vert_{L^2}}
{\bigl(\Vert e\Vert_{L^2}\,|e|_{H^2}\bigr)^{1/2}} \;\le\; 1
\qquad\text{on }\reals^{N_x}.
\label{eq:gap_tightness}
\end{equation}
The ratio $\rho_e$ measures the spectral \emph{bandwidth} of the error,
not its size, with $\rho_e\to1$ for a narrow-band error concentrated near a single
characteristic wavenumber
$k_e=\Vert\nabla e\Vert_{L^2}/\Vert e\Vert_{L^2}$, and $\rho_e\ll1$ for
broadband error.  If, and only if, fit errors are effectively narrow-band,
the gradient error is the geometric mean of the value
error and the curvature error, and
Eq.~\eqref{eq:gap_interpolation} turns from a bound into a
\emph{prediction}.

\begin{figure*}[t]
\centering
\includegraphics[width=0.96\textwidth]{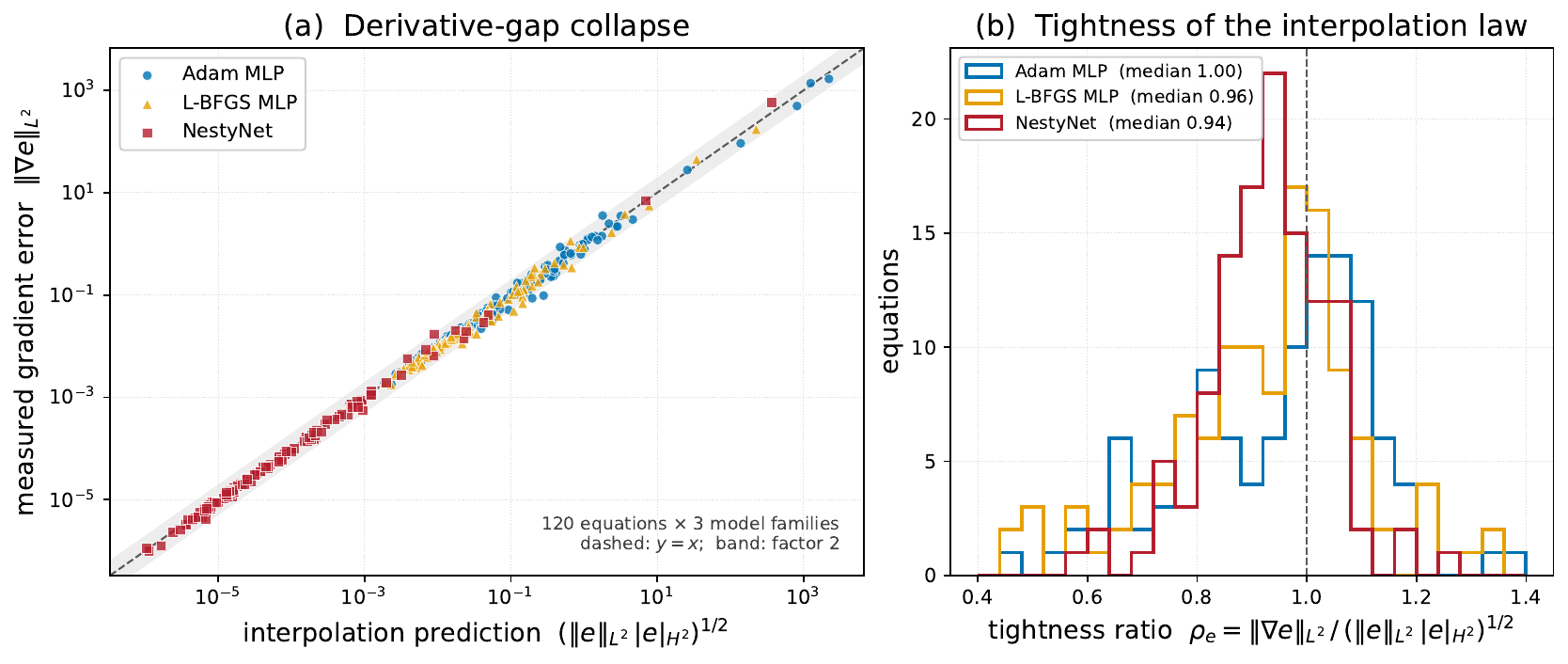}
\caption{The derivative-gap interpolation law across the AI Feynman suite.
Panel (a): measured gradient error $\Vert\nabla e\Vert_{L^2}$ of the fit
error $e=\hat f-f$ against the interpolation prediction
$(\Vert e\Vert_{L^2}\,|e|_{H^2})^{1/2}$ of
Proposition~\ref{prop:derivative_gap}, for all 120 equations and all
three model families (Adam MLPs, circles; L-BFGS-refined MLPs, triangles;
NestyNet, squares).  The dashed line marks equality and
the grey band a factor of two.  The 360 fits span more than nine orders of
magnitude in gradient error, and 356 lie within the factor-two
band.  Panel (b): distribution of the tightness ratio $\rho_e$ of
Eq.~\eqref{eq:gap_tightness} per model family; medians are
$0.94$--$1.00$, and $91.4\%$ of fits lie within a factor of $1.5$ of the
law.}
\label{fig:derivative_gap_collapse}
\end{figure*}

Figure~\ref{fig:derivative_gap_collapse} tests this prediction on every
equation of the benchmark and all three model families, evaluating the
norms as root-mean-square averages over the same probe samples as
Figure~\ref{fig:adam_vs_nestynet_scatter}, with the full Hessian in
Frobenius norm.  (On the bounded benchmark domains, sampled with the
SRBench measure, Proposition~\ref{prop:derivative_gap} holds only up to
boundary terms and sampling fluctuations, so ratios slightly above unity
occur).  The measured gradient errors collapse onto the interpolation
prediction over more than nine orders of magnitude. The median tightness
ratio is $0.94$--$1.00$ per model family, $356$ of $360$ fits lie within a
factor of $2$ of the prediction, and $91.4\%$ lie within a factor of
$1.5$.  The fitted errors in this benchmark behave as if they are effectively
narrow-band.
Training resolves the target up to a moving resolution frontier, and the
residual error concentrates near a characteristic wavenumber.  We also
see that the law is model-agnostic, given that MLPs and segmented models 
lie on the same line, and what distinguishes them is where along the line they sit.

Three consequences follow.  First, the regularity of the improvement
factors in Figure~\ref{fig:adam_vs_nestynet_scatter} is explained
quantitatively.  In the RMS norms used here, the median Adam-to-NestyNet
improvement factors are $2\,016\times$ for values, $960\times$ for curvature,
and $1\,466\times$ for gradients (these differ slightly from the per-axis
medians quoted for Figure~\ref{fig:adam_vs_nestynet_scatter} because the
norms here aggregate all axes and the full Hessian).  
Evaluated per equation (not by combining the three
aggregate medians above, which need not satisfy the identity term by
term), the geometric mean of the value and curvature gains predicts the
gradient gain at $1\,382\times$ against the measured $1\,466\times$, with
a median absolute offset of $16\%$.  For the L-BFGS
baselines the prediction is $438\times$ against a measured $422\times$
(median absolute offset $12\%$).  The multiplicative coupling can be understood as follows.
Because the gradient gain is the geometric mean of the value and
curvature gains, improving the value fit while letting the curvature
error grow buys almost nothing in gradients, and value accuracy converts
into gradient accuracy at an exchange rate set by curvature control.
Fitting values buys gradient accuracy exactly to the extent that
curvature error is simultaneously controlled.

Second, the law supplies an inexpensive a posteriori \emph{upper bound}
on the gradient error that requires no derivative data.
The value error $\Vert e\Vert_{L^2}$ is measurable on held-out function values, while the
curvature factor is bounded by
$|e|_{H^2}\le|\hat f|_{H^2}+|f|_{H^2}$, in which the model term is
available analytically.  It becomes a numerical estimate only on the
additional empirical premise that the error is narrow-band, $\rho_e\approx1$.
The two statements should not be conflated, and so in \S\ref{sec:cbe_application}
we report $\rho_e$ rather than assuming it.  Moreover, from Eq.~\eqref{eq:hessian}, using
$g_{os}(1-g_{os})\le\tfrac14$ and
$\Vert K_{os}K_{os}^\top\Vert_F=\Vert K_{os}\Vert_2^2$, the segmented
architecture carries a closed-form pointwise curvature budget,
\begin{equation}
\biggl(\sum_{ij}
\Bigl(\frac{\partial^2 \hat f_o}{\partial x_i\,\partial x_j}\Bigr)^{\!2}
\biggr)^{\!1/2}
\;\le\;
\frac{1}{4}\sum_{s}|a_{os}|\,\Vert K_{os}\Vert_2^{2}.
\label{eq:curvature_budget}
\end{equation}
Given an estimate of the target's curvature norm, often available from
the physics of the problem, the gradient error of a value-trained fit is
then bounded without differentiating any data.  And because
$|e|_{H^2}$ coincides with the $L^2$ error of the Laplacian, the same
factor governs the Laplacian part of the PDE residuals of
\S\ref{sec:applications}.

Third, the curvature factor is where the architectures separate.
Comparing each model's own curvature norm with the target's, the
segmented fits track the true curvature with median
$|\hat f|_{H^2}/|f|_{H^2}=1.000$ (5--95\% range $0.999$--$1.000$).
For 119 of the 120 equations this ratio remains below $1.1$; the sole
outlier reaches $1.17$.  Thus the canonical
initialization and residual-aligned segment growth usually add curvature
only where the residual demands it.  The Adam-trained MLPs show a median of
$0.99$ but a 5--95\% range of $0.82$--$1.60$, over-curving by more than $10\%$ on
$25$ of the $120$ equations (by up to $3.31\times$).  Through
Eq.~\eqref{eq:gap_interpolation}, excess
curvature of this kind is precisely what corrupts gradients even when
the values are well fit.

\subsection{Computational Speed}
\label{sec:comp_speed}

\begin{figure*}[t]
\centering
\includegraphics[width=0.48\textwidth]{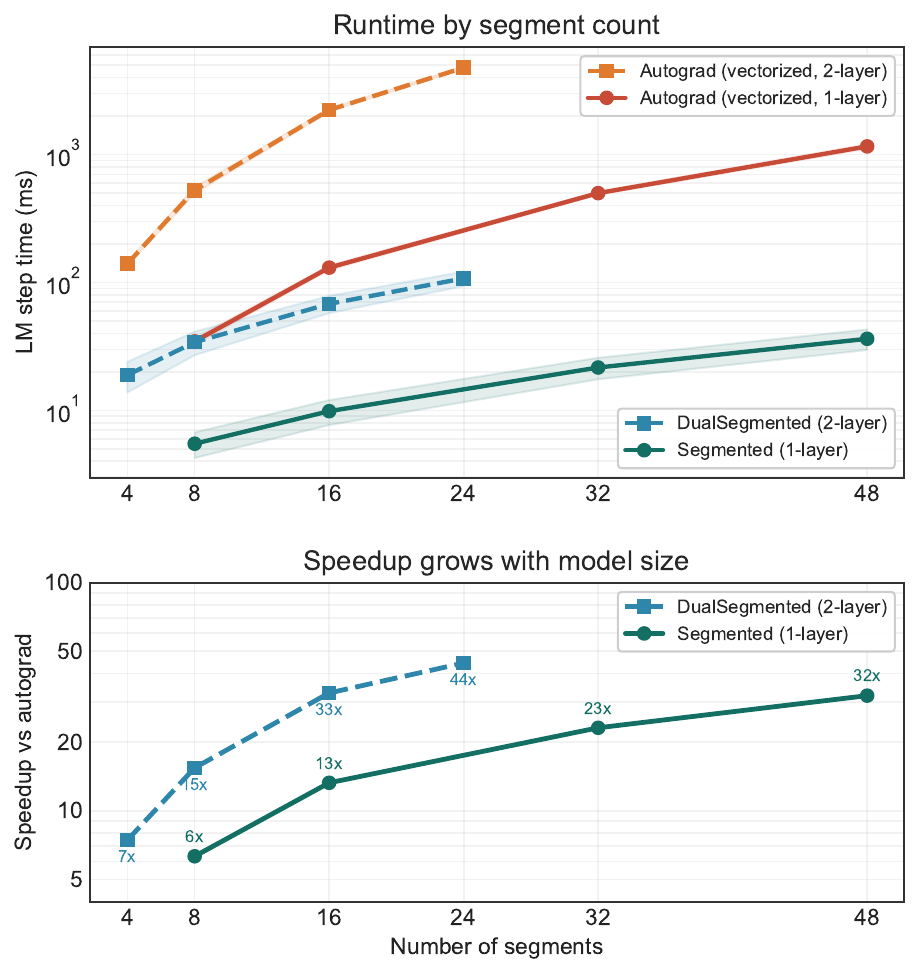}
\hfill
\includegraphics[width=0.48\textwidth]{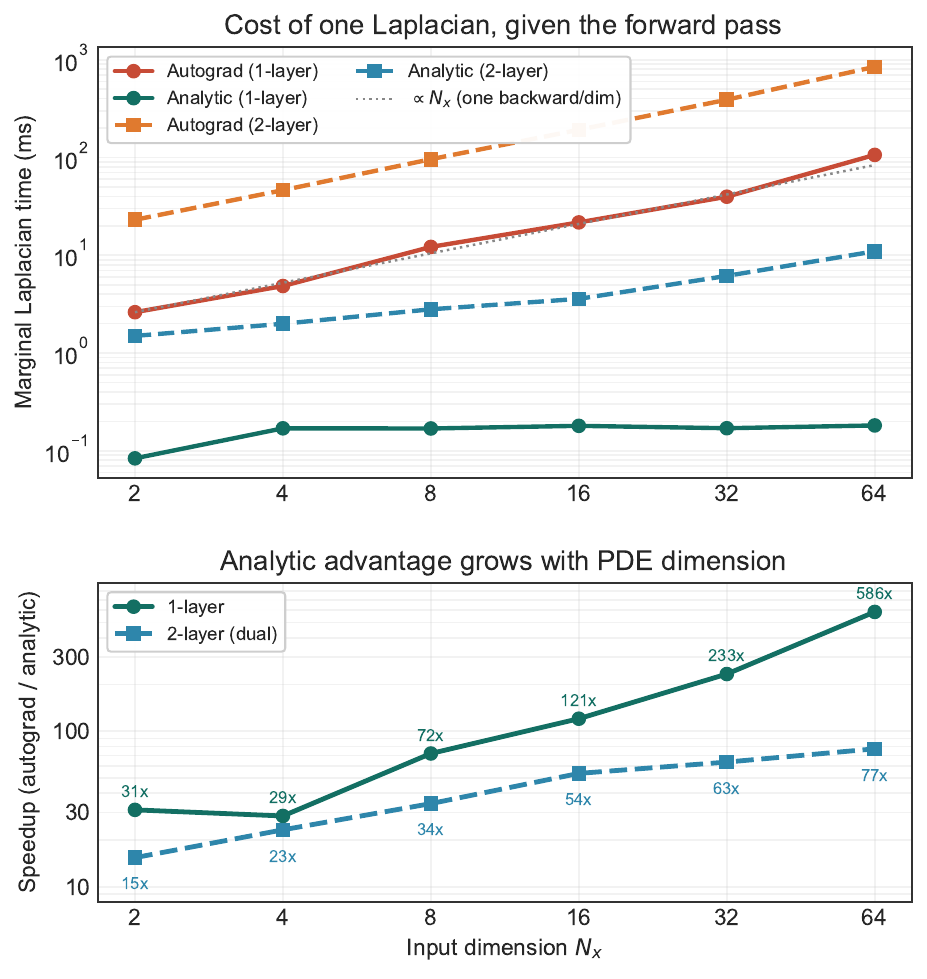}
\caption{Analytic versus autograd cost, end to end.  \emph{Left:} LM-step
runtime versus segment count for one-layer segmented and two-layer
dual-segmented models fit to AI Feynman problem \#29 (top), and the speedup
relative to a vectorized functorch \texttt{jacfwd} autograd baseline
(bottom).  The analytic advantage grows with model size, from
$\approx 6\times$ to $\approx 32\times$ (one-layer, $120$--$720$ parameters)
and $\approx 7\times$ to $\approx 44\times$ (two-layer, $240$--$1\,440$
parameters). Matrix-free analytic products cost linearly in $P$, while
autograd materializes the $R\times P$ Jacobian and its $P\times P$ normal
product.  \emph{Right:} Laplacian cost, analytic (cached) versus autograd,
versus input dimension $N_x$ at fixed architecture: marginal time (top) and
speedup (bottom).  Autograd rises linearly in $N_x$ (one backward pass per
dimension) while the analytic cached trace stays flat (one-layer) or gently
rising (two-layer). The speedup values are quoted in the text.  Analytic
and autograd compute identical operators to floating-point round-off.  All
timings: single CPU thread, double precision, full batch of
$2\times10^{3}$ points.}
\label{fig:segmented_dual_vs_autograd_speedup}
\label{fig:pinn_laplacian_scaling}
\end{figure*}

A natural concern is whether analytic derivative formulas come at the price
of slower evaluation.  The decisive test is whether the analytic
advantage in the derivative primitives (the Jacobian-vector and
vector-Jacobian products that the optimizer
composes into gradients and Gauss--Newton systems) survives once a full
Levenberg--Marquardt step is timed end to end.
The left panels of Figure~\ref{fig:segmented_dual_vs_autograd_speedup}
answer this directly.

The analytic path is dramatically faster.  Once residual evaluation,
derivative propagation, and step assembly are all included, the analytic
segmented models run tens of times faster than a vectorized autograd
baseline%
\footnote{All timings are single-CPU (Apple~M4, no GPU), double
precision, PyTorch~2.7.1, full-batch on $2\times10^{3}$ points, and are
steady-state, with three independent repeats after one warm-up step, which absorbs
one-time costs such as kernel and algorithm selection and functorch tracing.
The autograd baseline is the fastest \emph{correct} functorch route (a
vectorized forward-mode \texttt{jacfwd} Jacobian rather than a per-column loop).  
Neither path uses \texttt{torch.compile}, as we
noticed that it silently miscompiles the forward-mode autograd Jacobian to zero on
this workload, so eager vectorization is the fastest correct autograd.  Both
routes evaluate the identical derivative object, agreeing to floating-point
round-off (verified by the released derivative-audit tests).}, and the
margin widens as the models grow, reaching $\approx 32\times$ (one-layer)
and $\approx 44\times$ (two-layer) at the largest sizes tested.  The
mechanism is a difference in scaling.
The analytic path avoids materializing the full $R\times P$ autograd
Jacobian ($R$ residual rows, $P$ parameters) and supplies matrix-free
Jacobian and Gauss--Newton products at cost
linear in the number of active hinge atoms, whereas the autograd path forms
that $R\times P$ Jacobian and its $P\times P$ product and grows as $P^{2}$.
The speedup therefore climbs roughly linearly with model size.

\subsection{Input-Space Derivatives and Physics-Informed Networks}
\label{sec:pinn_speed}

The comparison above times derivatives with respect to \emph{parameters}, the
objects a fitting optimizer consumes.  A physics-informed network
(PINN)~\citep{Raissi2019,Karniadakis2021} instead consumes derivatives with
respect to the \emph{inputs} (gradients, Hessians, and above all the
Laplacian $\Delta f=\sum_i \partial^2 f/\partial x_i^2$) re-evaluated at
collocation points on every training step, so the analytic design has a
structural advantage worth quantifying separately.

NestyNet builds a derivative-enabled cache once per collocation batch (the
same forward pass the residual value already needs) and extracts the
Laplacian through a closed-form trace that never instantiates the dense
$N_x\times N_x$ Hessian.  Reverse-mode autograd (the backpropagation used to train neural
networks) instead computes the Laplacian
with one backward pass per input dimension, so its cost is $\mathcal{O}(N_x)$
passes, the familiar bottleneck of PINN training, whereas the analytic trace
is $\mathcal{O}(1)$ in passes.  For the two fitted models of
Figure~\ref{fig:segmented_dual_vs_autograd_speedup} at their native $N_x=3$,
the marginal cost of $\Delta f$ (the extra work once the forward/cache
exists, the operative quantity in a residual that also needs $f$) is $0.49$
versus $7.8$\,ms for the one-layer model and $3.4$ versus $16.2$\,ms for
the two-layer model, analytic versus autograd ($15.9\times$ and
$4.8\times$, or $2.7\times$ and $2.3\times$ for a lone Laplacian
built from scratch).  The analytic and autograd Laplacians agree
to floating-point round-off ($\le 4\times10^{-9}$), so this is a pure cost
comparison of identical operators.

This modest $N_x=3$ margin reflects an autograd penalty of only three
backward passes.  Its true size is a matter of
scaling.  The right panels of Figure~\ref{fig:pinn_laplacian_scaling} sweep the input
dimension with the architecture held fixed. The autograd Laplacian rises
linearly in $N_x$ while the analytic marginal cost stays essentially flat for
the one-layer model (and rises only gently for the two-layer model, whose
composite trace carries an $\mathcal{O}(N_x)$ Jacobian contraction).  The speedup
therefore climbs from $\approx 30\times$ at $N_x=2$ to $\approx 590\times$ at
$N_x=64$ for the one-layer model, and from $\approx 15\times$ to
$\approx 77\times$ for the two-layer model.  For the low-dimensional PDEs of
this paper the gain is a useful constant, but for the high-dimensional problems
where PINNs are most needed (Fokker--Planck, Schr\"odinger, or
Hamilton--Jacobi--Bellman equations in tens of variables) it is one to nearly
three orders of magnitude, and it is exact.

\subsection{Matrix-Free SPLA at Large Parameter Counts}
\label{sec:spla_performance}

This subsection previews what the segment-native preconditioner of
\S\ref{sec:spla} delivers once models grow beyond the dense direct-solve regime
of the benchmarks above.

\begin{figure}[t]
\centering
\includegraphics[width=0.98\columnwidth]{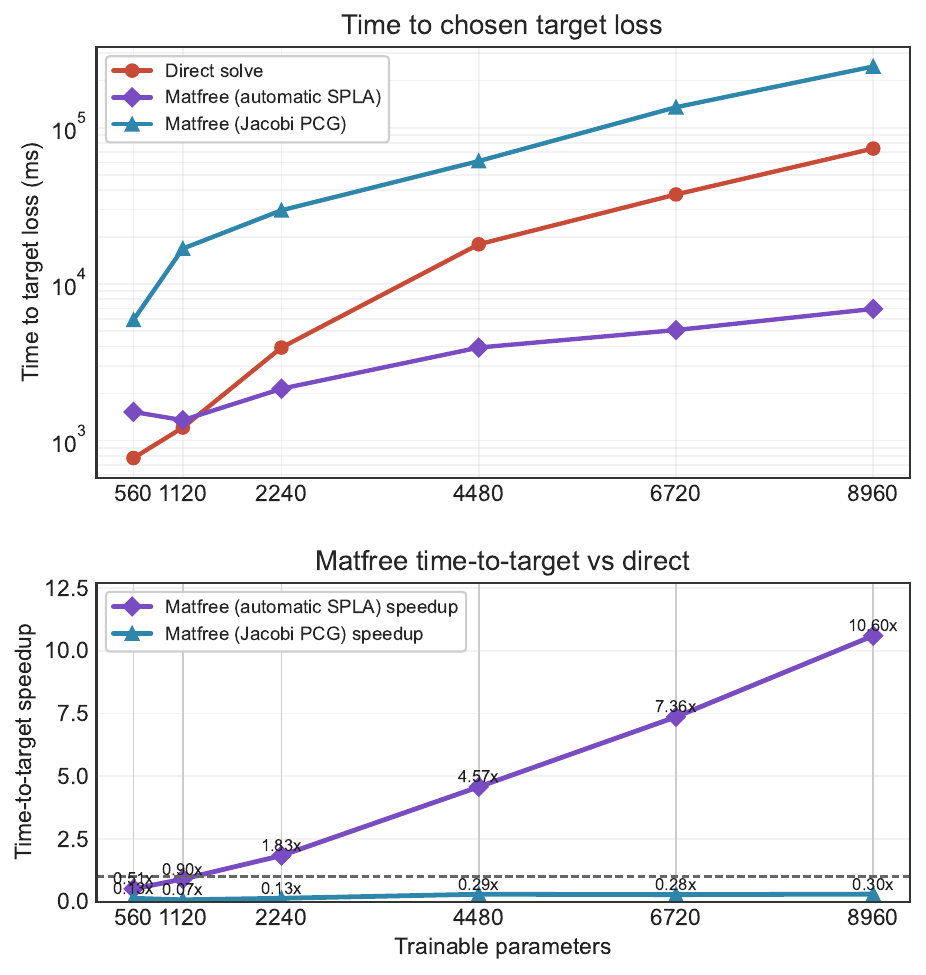}
\caption{End-to-end runtime comparison of three inner LM solvers on AI
Feynman problem \#11, the five-input target
$x_0\,[x_1+x_2x_3\sin x_4]$, as the single-layer segment count grows.  Each solver
is run from identical initializations to the same training loss of
$10^{-3}$. The top panel shows wall-clock time to target (log scale) and the
bottom panel the speedup of automatic SPLA over the dense direct solve.  The
three solvers are the dense direct solve (Cholesky, with a robust PyTorch
\texttt{linalg.solve()} fallback), matrix-free PCG with automatic
SPLA, and matrix-free PCG with
only the diagonal (Jacobi) preconditioner.  Automatic SPLA holds its cost
nearly flat ($1.35$--$6.9$\,s across a $16\times$ range of parameter counts)
while the direct solve grows cubically. SPLA becomes faster than the direct
solve above $\approx1\,200$ parameters and reaches $10.6\times$ at
$\approx9\,000$ parameters ($35\times$ over Jacobi-preconditioned PCG there),
with the advantage still widening at the largest size tested.}
\label{fig:direct_vs_matfree_speedup}
\end{figure}

In Figure~\ref{fig:direct_vs_matfree_speedup} we time three inner solvers
end-to-end on a representative AI Feynman fit (problem~\#11, the
five-input target $x_0\,[x_1+x_2x_3\sin x_4]$), measuring the
wall-clock time to reach a training loss of $10^{-3}$ as the parameter count
grows: the dense direct solve, matrix-free PCG with automatic SPLA (whose
automatically-selected recipe for this problem is the adaptive spectral Nystr\"om level of
\S\ref{sec:spla}), and matrix-free PCG with only the diagonal (Jacobi)
preconditioner.  For the
smallest models the dense direct solve is fastest, as its cubic
factorization cost is then negligible and it pays none of the
per-iteration cost of PCG (a matrix--vector product plus a preconditioner
application at every iteration).
But because the preconditioned matrix-free solver holds its iteration count
nearly fixed (and because the spectral level tunes its sketch rank to the
operator's effective dimension rather than to $n$), its per-step cost grows
only linearly with $n$, while the dense factorization grows cubically.  Both
solvers run the same predictive-LM trajectory and accept essentially the
same number of steps at every size, so their wall-clock ratio isolates the
inner linear solve. The two curves cross near $1\,200$ parameters, and the
advantage is still widening at the largest size tested (see caption).  The
near-unit preconditioned condition numbers
behind this flat scaling, and the theory that explains them, are
quantified in the SPLA theory paper.

\subsection{Predictive Mechanism Effectiveness}
\label{sec:predictor_ablation}

In the dense direct-solve regime used throughout these benchmarks, the
relevant predictive components are linear refinement and quadratic path
extrapolation.

Table~\ref{tab:direct_predictors_pb011} shows the direct-solve ablation on
AI Feynman problem~\#11.  Linear refinement has a large early effect.
After 100 LM steps, the validation loss drops from $6.6\times 10^{1}$ in
the baseline run to $2.4\times 10^{-3}$.  Thus linear refinement is
primarily an early-basin accelerator in this experiment.  Quadratic path
extrapolation then extends the run to a lower final validation loss,
accepting 329 extrapolated steps and reaching $4.4\times 10^{-6}$, so it
improves the final attainable fit on this problem.  This is a single
example, but here linear refinement drives the early basin and 
quadratic extrapolation the endgame.  Whether the two mechanisms 
help, and in which phase, is problem-dependent (\S\ref{sec:predictive}).

\begin{table}[htbp]
\centering
\small
\setlength{\tabcolsep}{3pt}
\begin{tabular}{lrrrrrr}
\toprule
\textbf{Config.} &
\textbf{Steps} &
\shortstack{\textbf{Wall}\\\textbf{(s)}} &
\shortstack{\textbf{Val.}\\\textbf{iter. 100}} &
\shortstack{\textbf{Val.}\\\textbf{final}\\{\scriptsize$\times10^{-5}$}} &
\shortstack{\textbf{Lin.}\\\textbf{ref.}} &
\shortstack{\textbf{Quad.}\\\textbf{acc.}} \\
\midrule
Baseline & $1\,601$ & 83.7  & $66$     & $1.1$  & 0 & 0 \\
+ Linear & $2\,301$ & 180.3 & $0.0024$ & $1.4$  & 522 & 0 \\
+ Quad.  & $4\,401$ & 248.8 & $0.0024$ & $0.44$ & 868 & 329 \\
\bottomrule
\end{tabular}
\caption{Dense direct-solve ablation on AI Feynman problem \#11.  ``Val.\,iter. 100'' is the
validation loss after 100 LM steps and ``Val.\ final'' the final validation loss (in units of
$10^{-5}$); ``Linear'' and ``Quad.'' count
linear-refinement applications and quadratic-extrapolation accepts,
respectively.  Linear refinement provides a large early convergence gain,
while quadratic path extrapolation improves the final attainable validation
loss in this run.}
\label{tab:direct_predictors_pb011}
\end{table}

\section{An Astrophysical Application: Vertical Accelerations in Galactic Disks}
\label{sec:cbe_application}

Papers~II--IV build inference and discovery machinery on the NestyNet substrate.  Our
aim in this section is narrower, to check that the interpolation law of
\S\ref{sec:derivative_gap} is usable in a physical inverse problem, where the
deliverable is a derivative rather than a value.  The vertical acceleration
$K_z(z)$ of the Galactic disk is a natural test.  Its
measurement is a century-old problem in Galactic
dynamics~\citep{Kapteyn1922,Oort1932,Bahcall1984,KuijkenGilmore1989,KuijkenGilmore1991,Creze1998,HolmbergFlynn2000},
renewed by modern spectroscopic and astrometric surveys as a probe of the
local dark-matter
density~\citep{BovyTremaine2012,Zhang2013,BovyRix2013,Bienayme2014,Read2014,Widmark2019,Salomon2020},
and its observable is a ratio of derivatives, so its accuracy is governed by
exactly the gradient error that Proposition~\ref{prop:derivative_gap} bounds.
For a stationary,
phase-mixed stellar population the collisionless Boltzmann equation
reduces~\citep{BinneyTremaine2008}, in the vertical phase space $(z,v)$, to
\begin{equation}
v\,\partial_z \ln f + K_z(z)\,\partial_v \ln f = 0,
\label{eq:cbe}
\end{equation}
so the acceleration is fixed by the log-gradients of the distribution
function $f(z,v)$ alone. The overall normalization of $f$ cancels, so no
stellar \emph{density} is required.  
We fit the Poisson-weighted binned star counts with $\hat g=\ln\hat f$.
Because Eq.~\eqref{eq:cbe} is linear in $K_z$, we recover it by
least squares in each $z$-bin,
\begin{equation}
\hat K_z(z_j)=-\,\frac{\sum_{i} v_i\,\hat g_{z,i}\,\hat g_{v,i}}
{\sum_{i}\hat g_{v,i}^{2}},
\qquad \hat g_z=\partial_z\hat g,\quad \hat g_v=\partial_v\hat g,
\label{eq:cbe_gls}
\end{equation}
from the analytic gradients of \S\ref{sec:derivatives} (the global
solve avoids the ill-conditioned pointwise ratio near zero vertical velocity,
where $\hat g_v\to0$.  For the isothermal slab $g_v=-v/\sigma^2$, so this occurs
at $v=0$ at every height, whereas at the midplane it is $g_z$ and $K_z$ that
vanish).  Equation~\eqref{eq:cbe_gls} is the unweighted normal equation.  Our
implementation adds a Tikhonov regularization ridge and, when the differentiated-CBE rows
of \S\ref{sec:cbe_hierarchy} are included, a relative weight between the
two row blocks.

We test the estimator on a mock self-gravitating isothermal slab, whose
potential $\Phi(z)=2\sigma^2\ln\cosh(z/a)$ gives the closed-form truth
$K_z(z)=-(2\sigma^2/a)\tanh(z/a)$.  From $5\times10^{5}$ stars drawn from the
stationary distribution $f\propto e^{-(v^2/2+\Phi)/\sigma^2}$, binned on a
$48\times48$ grid in $(z,v)$, we fit $\hat g$ to the Poisson-weighted
log-counts (with the small-count correction
$\mathbb{E}[\ln N]\simeq\ln\mu-\tfrac1{2\mu}$) and solve
Eq.~\eqref{eq:cbe_gls} in $10$ bins across the well-sampled core $|z|\le a$.
The statistical uncertainty follows from disjoint split-halves of the
catalog, and so remains available on real data, where the true $K_z$ is
unknown.

\begin{figure*}[t]
\centering
\includegraphics[width=\textwidth]{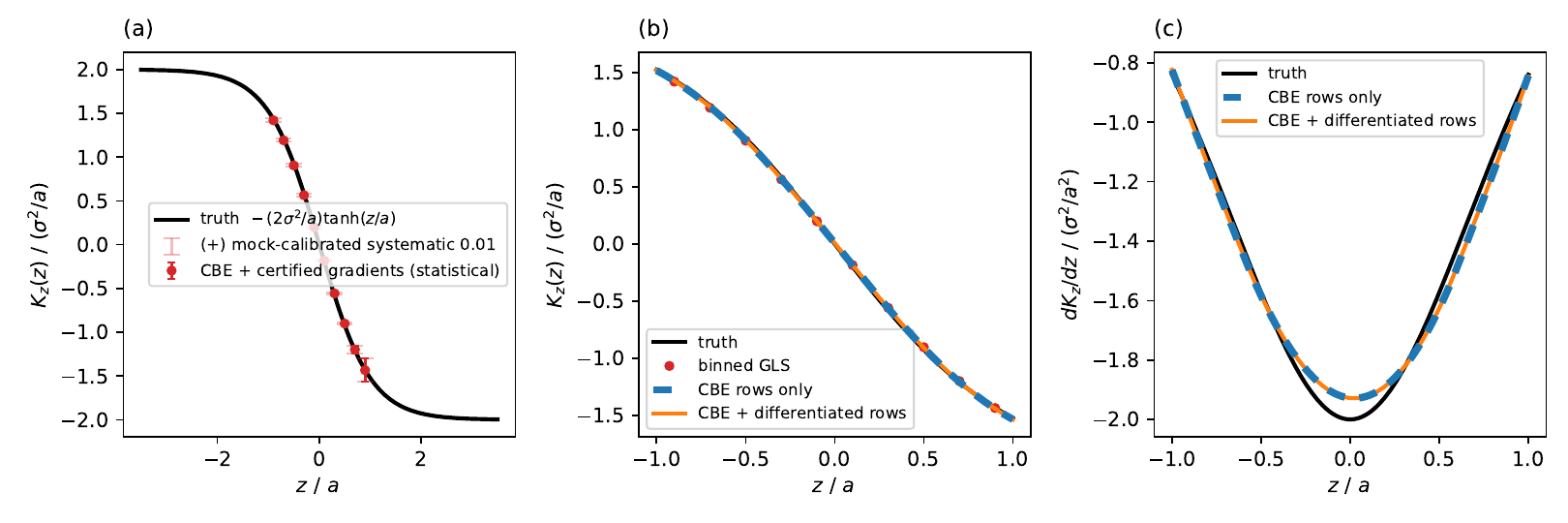}
\caption{Statistical recovery of the Galactic vertical acceleration and its
derivative tower on a mock isothermal slab.  (a)~The certified-gradient
estimator of Eq.~\eqref{eq:cbe_gls} in $10$ bins across the core $|z|\le a$,
from $5\times10^5$ stars.  The inner bars are the split-half statistical
uncertainty, which needs no knowledge of the truth, and outer bars add the
mock-calibrated systematic, which does.  The solid curve is the analytic
truth $K_z=-(2\sigma^2/a)\tanh(z/a)$, and the segment count is
selected by the derivative-space capacity rule of the text, not
tuned.  (b)~The same force from the smooth coefficient field
$\sum_\ell\beta_\ell B_\ell(z)$, solved from the CBE rows alone and with
the differentiated-CBE rows added, against the binned estimator (points).  (c)~The
density-sensitive derivative $K_z'(z)=\sum_\ell\beta_\ell B_\ell'(z)$,
obtained analytically from the same coefficients.  Through the slab Poisson
relation this is the local mass density up to the constant $-1/(4\pi G)$.}
\label{fig:kz_recovery}
\label{fig:kz_tower}
\end{figure*}

Proposition~\ref{prop:derivative_gap} allows us to \emph{select the model capacity}.
Among fits that are adequate in value, $\Vert e\Vert_{L^2}$ is pinned at the
Poisson noise floor, so by Eq.~\eqref{eq:gap_interpolation} the gradient error
grows with any excess curvature the fit injects.  The derivative-optimal
segment count is thus the smallest that remains value-adequate, and it can be
identified from the data alone by minimizing a held-out residual of
Eq.~\eqref{eq:cbe} in place of a value criterion.  Here it selects a
four-segment model.  Figure~\ref{fig:kz_recovery}a shows the
recovered acceleration, which matches the truth to a
root-mean-square error of $\approx 0.015\,\sigma^2/a$ across $|z|\le a$,
against a signal of order $\sigma^2/a$, and the tightness ratio of
Eq.~\eqref{eq:gap_tightness} holds at $\rho_e\approx1$ as the fitted subsample
grows from $2.5\times10^4$ to $4\times10^5$ stars. This is the a posteriori gradient-error
estimate of Eq.~\eqref{eq:gap_interpolation}, realized on a physical inverse
problem.  A full survey measurement (measurement-error deconvolution, spatial
selection gradients, and departures from stationarity, for which the residual
of Eq.~\eqref{eq:cbe} is itself the diagnostic) is left to future work.  The interesting
point here is that certified derivatives are able to turn a ratio of noisy gradients into
a calibrated physical observable.

\subsection{From a Single Tracer to a Hierarchy}
\label{sec:cbe_hierarchy}

The vertical-force experiment is the scalar, single-population, first-derivative
rung of a broader hierarchy, and two extensions are worth
making explicit.  The first is a \emph{derivative tower}.  The Fourier
Cauchy--Schwarz argument behind Proposition~\ref{prop:derivative_gap} iterates,
$\Vert|\nabla|^{m}e\Vert_{2}^{2}\le\Vert|\nabla|^{m-1}e\Vert_{2}\,
\Vert|\nabla|^{m+1}e\Vert_{2}$, so the derivative-error sequence is log-convex
and, for narrow-band error, geometric.  On bounded and sampled data this is an
empirical diagnostic certificate, but it ties each
deliverable to the next.  This matters for our Galactic dynamics example because the mass density needs one
derivative more than the force.  Representing $K_z$ in a smooth basis
$K_z(z)=\sum_\ell\beta_\ell B_\ell(z)$ keeps Eq.~\eqref{eq:cbe} linear in the
coefficients, $r_i^{(0)}=v_i\hat g_{z,i}+\hat g_{v,i}\sum_\ell\beta_\ell
B_\ell(z_i)$, and the density-sensitive derivative $K_z'$ can be constrained
by appending the $z$-differentiated CBE as a second row block, still
linear in $\beta$ and consuming only the analytic second derivatives
$\hat g_{zz}$ and $\hat g_{zv}$.  The slab Poisson equation then
gives $\rho(z)\simeq-K_z'(z)/(4\pi G)$ (with a radial term in a fully
axisymmetric analysis), so the same interpolation law that selects a
derivative-safe model for $K_z$ also supplies the capacity diagnostic for the
next deliverable, the local mass density.  Figures~\ref{fig:kz_tower}b
and \ref{fig:kz_tower}c show both rungs recovered from a single fit.

\begin{figure}[tb]
\centering
\includegraphics[width=\columnwidth]{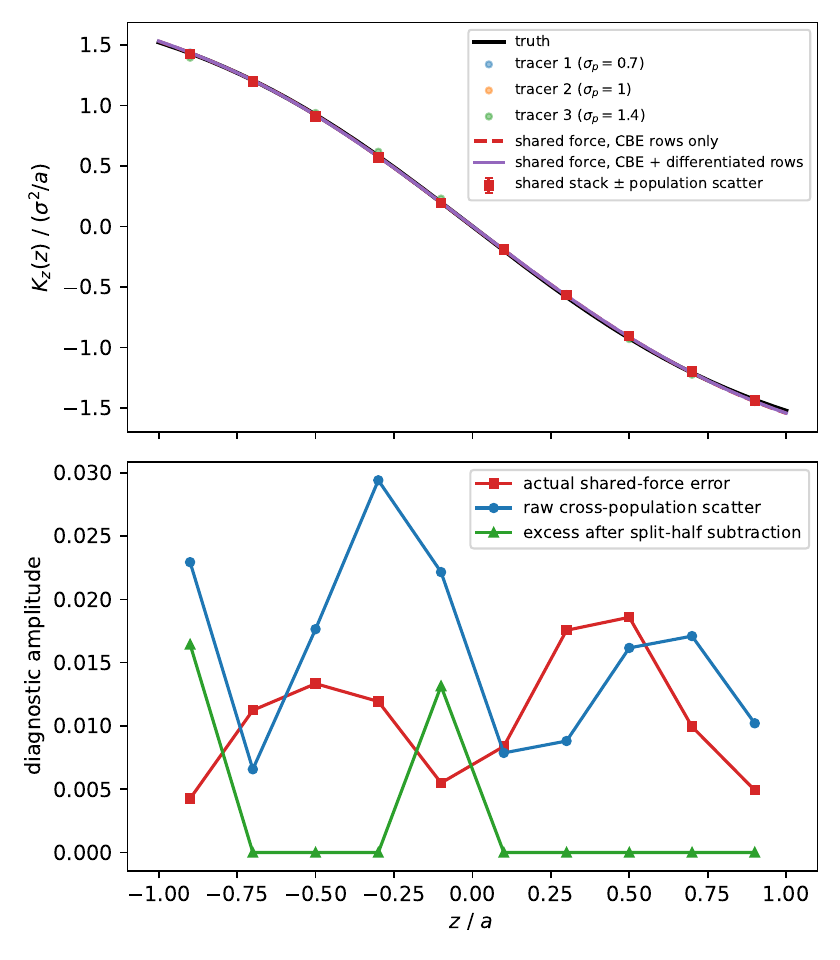}
\caption{The population hierarchy.  Top: three mock tracer populations of
dispersion $\sigma_p=0.7,1,1.4$ (in units of the potential's $\sigma$) in one
shared potential, each recovering $K_z(z)$, with the stacked shared-force
solution (squares, error bars showing the cross-population scatter).  Bottom:
the per-bin diagnostics, the actual shared-force error, the raw cross-population
scatter, and the \emph{excess} scatter after subtracting the split-half
statistical variance.  The excess is a data-only estimate of population-dependent model-class
systematics.}
\label{fig:kz_population}
\end{figure}

A second hierarchy is supplied by the data themselves.  Distinct stellar
populations, sliced by chemistry or age, have different distribution functions
but share the same potential, so one may fit a separate $\hat g_p=\ln\hat f_p$
per population while solving a single \emph{stacked} least-squares problem
$r_{ip}=v_i\partial_z\hat g_p+K_z(z_i)\partial_v\hat g_p$ for the common
force $K_z$.
Each tracer projects its own $\ln f_p$ onto the model class with its own error,
while $K_z$ is shared.  The excess cross-population scatter of the recovered
force, after subtracting the split-half statistical variance, is therefore an
empirical, per-bin estimate of population-dependent model-class systematics,
computed with no knowledge of the truth.  It does not capture common-mode
errors, such as a shared selection gradient or a bias common to every
fit.  Two further checks can cover those.  The first is calibration on mocks
matched to the survey, as in Figure~\ref{fig:kz_recovery}a.  The second
is parity. For a stationary population in a midplane-symmetric disk the
distribution function is even under $(z,v)\to(-z,-v)$, so any odd
component of the fit is a data-only diagnostic of asymmetric systematics
or disequilibrium.  What the cross-population scatter adds is a component
of the systematics budget that no single tracer can supply.
Figure~\ref{fig:kz_population} shows the stacked recovery together with
the per-bin diagnostic, whose excess concentrates in the bins where the
coldest tracer is sparsest and its fit degrades.

The natural next step would extend the same construction to axisymmetric
phase space, where the CBE becomes a varying-coefficient identification
of the acceleration field $(a_R,a_z)(R,z)$, overdetermined across
velocity space at each point.  One could fit the field directly, which
keeps the estimator a linear generalized least-squares solve, or fit a
scalar potential instead, trading that linearity for exact
curl-free solutions.  Departures from stationarity would then appear as the
residual of Eq.~\eqref{eq:cbe}, with the
phase spiral~\citep{Antoja2018} being the prime example of what such a
residual would contain.  In every case the estimator would remain the
same object as in the $K_z$ demonstration, a conservation law linear in
unknown coefficient fields, applied to certified derivatives of a fitted
phase-space surface.

\section{Conclusions}
\label{sec:conclusions}

We have presented NestyNet, a framework for neural network surrogates
aimed at scientific targets where high-fidelity representations induce strongly
correlated, stiff, and sloppy parameter directions. In this regime
standard MLP-plus-first-order training is inadequate and derivative
inaccuracy becomes the most visible symptom.  NestyNet addresses the
mismatch by coupling a segmented analytic representation to a predictive
second-order solver designed around the resulting linear-algebra structure.

The first conceptual contribution is the model itself, a segmented softplus
surrogate matched to multiscale and sharply structured but still
differentiable targets, with closed-form gradients, Hessians, and
Laplacians, and (one rung up the same polylogarithm ladder) closed-form
line and hyperrectangle antiderivatives.  This design admits a deterministic ``canonical'' initialization
which embeds the best affine fit exactly through opposed softplus hinges and
then grows segments by residual-aligned projections, yielding repeatable fits
and low-overlap starting points.  This frees the fits from random initialization, which is valuable
wherever reproducibility matters.

The second is the optimizer, a predictive Levenberg--Marquardt algorithm that exploits
the segment-native structure of the models, with
solved-geometry parameter scaling, four predictive mechanisms (post-step
linear refinement, quadratic path extrapolation, a direct predictor with
Woodbury fast-path, and a local manifold subspace solve), and a
quasi-Newton L-BFGS rescue scheme,
each specialized to a different phase of the solve.

The third is SPLA, a segment-native preconditioning framework
for the small-damping regime that dominates physics LM solves.  Every
result in this paper is small enough to be optimized without it, but
as we show in the end-to-end preview it accelerates the
matrix-free solves by up to $35\times$ at the largest model sizes we test
(Figure~\ref{fig:direct_vs_matfree_speedup}).  The segmented structure
therefore does not merely supply convenient modeling blocks.  Its
theory, constructions, and diagnostics are developed in the companion SPLA theory
paper.

The fourth is a law we stumbled upon empirically. Across all $360$ fits of
the adopted benchmark, three model families, and more than nine orders of
magnitude in error, we find that the gradient error is the geometric mean of the value
and curvature errors, the interpolation bound of
Proposition~\ref{prop:derivative_gap} holding near equality (median
tightness $0.94$--$1.00$, with $356$ of $360$ fits within a factor of
two).  The bound itself is elementary Fourier analysis.  What gives it
power is that its narrow-band premise is both controllable and checkable
for segmented fits. It is controllable because the canonical construction adds
curvature only where the residual demands it, so the fitted curvature
norms track those of the targets (median ratio $1.000$). It is checkable because
$\rho_e$ can be measured directly, on matched mocks or through held-out
residuals, rather than assumed.  Once both hold, an inequality becomes an instrument.
It selects the derivative-optimal capacity from the data alone, budgets
gradient error without any derivative data, and in
\S\ref{sec:cbe_application} turns a ratio of noisy gradients into a
calibrated physical observable.

Empirically, on the AI Feynman benchmark NestyNet achieves $\sim 2\,100\times$
better function accuracy, $\sim 1\,400\times$ first-derivative accuracy, and
$\sim 4\,800\times$ better integral accuracy than Adam-trained MLPs.  Even after
full-batch L-BFGS refinement initialized from the Adam checkpoints, the
corresponding median gains remain $\sim 250$--$710\times$, with
operator-level gains of roughly $780$--$1\,400\times$ over Adam and
$250$--$400\times$ after L-BFGS refinement for directional derivatives,
mixed partials, directional curvature, and Laplacians.  The analytic derivatives provide
substantial speed gains compared to automatic differentiation (in PyTorch),
with an end-to-end LM-step advantage that grows with model size, reaching
$\approx 44\times$ against a vectorized-autograd baseline.
All of these results were obtained with a single, fixed hyperparameter
configuration applied across the entire benchmark, so the gains reflect the
method's robustness rather than any per-problem tuning.

These results have implications across scientific computing.  Accurate
analytic Laplacians remove one important source of error in PINN-style
residuals, and the operator-focused benchmark
(Figure~\ref{fig:adam_vs_nestynet_operator_scatter}) shows the advantage
extends to the mixed partials and contracted Hessian probes that enter PDE
residuals.  Accurate $\nabla H$ likewise enables symplectic integration
without secular energy drift.  A composable adaptor library extends the
same optimizer to PINNs, variable projection, implicit Hamilton--Jacobi
constructions, and vector- and complex-valued surrogates, and once the surrogate's
local differential structure is reliable it becomes a trustworthy substrate
for the uncertainty quantification of Paper~II and the downstream
discovery tasks of Papers~III and~IV.
Complementing the two end-to-end PINNacle benchmark wins, the supervised case-by-case
study clears the reference accuracy on the remaining 20 problems, 19 on
every data seed and Poisson2d-C on four of five,
showing that representation is not the limiting factor at those benchmark
scales.  Its practical use in physics-informed recovery remains a separate
challenge.

The main limitation of the segmented architecture is that it trades generic
expressivity for structured smoothness and analytic tractability.  For
instance, a sum of soft hinges may require more parameters than a
conventional MLP for highly oscillatory targets.  Extending the
construction to convolutional or attention-based architectures also
remains open.  The optimizer, in contrast, is not tied to the segmented
family.  An autograd adaptor presents any PyTorch model through the same
provider interface, with the derivative products then supplied by
automatic differentiation in place of the closed forms, so the
predictive LM machinery applies to general architectures and is simply
at its fastest on segmented ones.  Indeed the autograd baseline of
\S\ref{sec:comp_speed} is driven by the same optimizer through exactly
this adaptor.  The SPLA method also relies
on useful structure in the whitened spectrum of the normal matrix. The
preconditioner accelerates the solve only when that spectrum is
compressible, or when the chosen coarse space captures the dominant
cross-segment modes, and it degrades to a diagnostic otherwise.  The
adaptive spectral level provides the required on-the-fly behavior in the
compressible regime.  
Nonlinear constraints are currently imposed by
construction (structural parameterizations, boundary-condition and
implicit-solve providers) or as weighted residual rows. A native
optimizer-level treatment of general nonlinear parameter constraints
remains a natural next step.

The present work has focused on physics applications that require small
networks of up to a few thousand parameters.  The matrix-free derivative
design coupled with the spectral SPLA level already demonstrates the
feasibility of the highly accurate LM Gauss--Newton algorithm at nearly
$10^{4}$ parameters, a full order of magnitude faster than the dense direct
solve (Figure~\ref{fig:direct_vs_matfree_speedup}).  A systematic treatment of
the large-model regime will be explored in a future contribution.

\begin{acknowledgments}
RI gratefully acknowledges funding in the initial stages of this project
from the European Research Council (ERC) under the European Union's Horizon
2020 research and innovation programme (grant agreement No. 834148). We
gratefully acknowledge the High Performance Computing center of the
Universit\'e de Strasbourg for a very generous time allocation and for
their support over the development of this project. RI 
thanks Alonso Ibata for highly stimulating discussions that led to the development
of several modules for stochastic calculus in the codebase.
\end{acknowledgments}

\software{NestyNet (this work;
\url{https://github.com/RodrigoIbata/NestyNet}),
PyTorch v2.7.1 \citep{Paszke2019},
NumPy \citep{Harris2020},
SciPy \citep{Virtanen2020},
Matplotlib \citep{Hunter2007},
scikit-learn \citep{Pedregosa2011}.}

\bibliographystyle{aasjournalv7}
\bibliography{nestynet_paper1}

\end{document}